\documentclass[10pt,journal,letterpaper]{IEEEtran}
\pdfoutput=1 

\usepackage{paper}      
\usepackage{mathdefs}   

\begin{document}

\title{On Characterizing the Local Pooling Factor of Greedy Maximal Scheduling in Random Graphs}

\author{%
  Jeffrey~Wildman,~\IEEEmembership{Member,~IEEE,} and~Steven~Weber,~\IEEEmembership{Senior~Member,~IEEE}%
  \thanks{J.~Wildman and S.~Weber are with the Department of Electrical and Computer Engineering, Drexel University, Philadelphia, PA, USA (email: jww33@drexel.edu; sweber@coe.drexel.edu).}%
  \thanks{Preliminary results were presented at the 2013 Allerton Conference on Communication, Control, and Computing \cite{WilWeb2013}.}%
}

\IEEEpubid{%
  \begin{minipage}[t]{0.85\textwidth}%
    \centering\copyright~2015 IEEE.
    Personal use of this material is permitted.
    Permission from IEEE must be obtained for all other uses, in any current or future media, including reprinting/republishing this material for advertising or promotional purposes, creating new collective works, for resale or redistribution to servers or lists, or reuse of any copyrighted component of this work in other works.%
  \end{minipage}%
}

\maketitle

\begin{abstract}
  The study of the optimality of low-complexity greedy scheduling techniques in wireless communications networks is a very complex problem.
  The Local Pooling (LoP) factor provides a single-parameter means of expressing the achievable capacity region (and optimality) of one such scheme, greedy maximal scheduling (GMS).
  The exact LoP factor for an arbitrary network graph is generally difficult to obtain, but may be evaluated or bounded based on the network graph's particular structure.
  In this paper, we provide rigorous characterizations of the LoP factor in large networks modeled as \Erdos-\Renyi (ER) and random geometric (RG) graphs under the primary interference model.
  We employ threshold functions to establish critical values for either the edge probability or communication radius to yield useful bounds on the range and expectation of the LoP factor as the network grows large.
  For sufficiently dense random graphs, we find that the LoP factor is between 1/2 and 2/3, while sufficiently sparse random graphs permit GMS optimality (the LoP factor is 1) with high probability.
  We then place LoP within a larger context of commonly studied random graph properties centered around connectedness.
  We observe that edge densities permitting connectivity generally admit cycle subgraphs which forms the basis for the LoP factor upper bound of 2/3.
  We conclude with simulations to explore the regime of small networks, which suggest the probability that an ER or RG graph satisfies LoP and is connected decays quickly in network size.
\end{abstract}


\section{Introduction}\label{sec:introduction}

\IEEEPARstart{T}{he} \emph{stability region} (or capacity region) of a queueing network is often defined as the set of exogenous traffic arrival rates for which a stabilizing scheduling policy exists.
A scheduling policy is \emph{optimal} if it stabilizes the network for the entire stability region.
In \cite{TasEph1992}, Tassiulas and Ephremides proved the optimality of the Maximum Weight Scheduling (MWS) policy, which prioritizes backlogged queues in the network.
However, for arbitrary communication networks and interference models, employing MWS incurs large computation and communication costs.
Under the assumption of graph-based networks with primary interference, the MWS policy simplifies to that of the Maximum Weighted Matching (MWM) problem, for which there are polynomial-time algorithms.

Greedy and heuristic scheduling can help reduce these operating costs further, usually at the expense of optimality.
The relative performance of these policies is often defined by their achievable fraction \(\gamma\) of the stability region.
For example, Sarkar and Kar \cite{SarKar2008} provide a \(\Oh{\Delta\log\Delta\log n}\)-time (where \(\Delta\) is the max degree of the network) scheduling policy that attains at least \(2/3\) of the stability region for tree graphs under primary interference.
Lin and Shroff \cite{LinShr2006} prove that a maximal scheduling policy on arbitrary graphs can do no worse than \(1/2\) of the stability region under primary interference.
Maximal matching policies can be implemented to run in \(\Oh{\log^2 n}\)-time \cite{IsrIta1986}.
Lin and Rasool \cite{LinRas2009} propose a constant, \(\Oh{1}\)-time algorithm that asymptotically achieves at least \(1/3\) of the stability region under primary interference.
This naturally leads to the question of whether or not greedy scheduling techniques may in fact be optimal (\(\gamma = 1\)).

\subsection{Related Work}\label{sec:related-work}
\IEEEpubidadjcol 

Sufficient conditions for the optimality of Greedy Maximal Scheduling (GMS) employed on a network graph \(G(V,E)\) were produced by Dimakis and Walrand \cite{DimWal2006} and called Local Pooling (LoP).
The GMS algorithm (called Longest Queue First, LQF \cite{DimWal2006}) consists of an iterated selection of links in order of decreasing queue lengths, subject to pair-wise interference constraints.
Computing whether or not an arbitrary graph \(G\) satisfies LoP consists of solving an exponential number of linear programs (LPs), one for each subset of links in \(G\).
Trees are an example of one class of graphs proved to satisfy LoP.
While LoP is necessary and sufficient under deterministic traffic processes, a full characterization of the graphs for which GMS is optimal under random arrivals is unknown.

The work by Birand \ea~\cite{BirChuRie2012} produced a simpler characterization of all LoP-satisfying graphs under primary interference using forbidden subgraphs on the graph topology.
Even more remarkably, they provide an \(\Oh{n}\)-time algorithm for computing whether or not a graph \(G\) satisfies LoP.
Concerning general interference models, the class of \emph{co-strongly perfect} interference graphs are shown to satisfy LoP conditions.
The definition of co-strongly perfect graphs is equated with the LoP conditions of Dimakis and Walrand \cite{DimWal2006}.
Additionally, both Joo \ea~\cite{JooLinShr2009-ToN} and Zussman \ea~\cite{ZusBrzMod2008} prove that GMS is optimal on tree graphs for \(k\)-hop interference models.

For graphs that do not satisfy local pooling, Joo \ea~\cite{JooLinShr2009-ToAC, JooLinShr2009-ToN} provide a generalization of LoP, called \(\sigma\)-LoP.
The LoP factor of a graph, \(\sigma\), is formulated from the original LPs of Dimakis and Walrand~\cite{DimWal2006}.
Joo \ea~\cite{JooLinShr2009-ToN} show that the LoP factor is in fact GMS's largest achievable uniform scaling \(\sigma = \gamma^*\) of the network's stability region.
Li \ea~\cite{LiBoyXia2011} generalize LoP further to that of \(\Sigma\)-LoP, which includes a per-link LoP factor \(\sigma_l\) that scales each dimension of \(\Lambda\) independently and recovers a superset of the provable GMS stability region under the single parameter LoP factor.

As mentioned, checking LoP conditions can be computationally prohibitive, particularly under arbitrary interference models.
Therefore, algorithms to easily estimate or bound \(\sigma\) and \(\sigma_l\) are of interest and immediate use in studying GMS stability.
Joo \ea~\cite{JooLinShr2009-ToN} provide a lower bound on \(\sigma\) by the inverse of the largest interference degree of a nested sequence of increasing subsets of links in \(G\), and provide an algorithm for computing the bound.
Li \ea~\cite{LiBoyXia2011} refine this algorithm to provide individual per-link bounds on \(\sigma_l\).
Under the primary interference model, Joo \ea~\cite{JooLinShr2009-ToAC} show that \(\Delta / (2\Delta-1)\) is a lower bound for \(\sigma\).
Leconte \ea~\cite{LecNiSri2011}, Li \ea~\cite{LiBoyXia2011}, and Birand \ea~\cite{BirChuRie2012} note that a lower bound for \(\sigma\) is derived from the ratio of the min- to max-cardinality maximal schedules.

Joo \ea~\cite{JooLinShr2009-ToN} define the worst-case LoP over a class of graphs, and in particular find bounds on the worst-case \(\sigma\) for geometric-unit-disk graphs with a \(k\)-distance interference model.
Birand \ea~\cite{BirChuRie2012} list particular topologies that admit arbitrarily low \(\sigma\), and provide upper and lower bounds on \(\sigma\) for several classes of interference graphs.
The body of work by Brzezinski \ea~\cite{BrzZusMod2008-ITA, ZusBrzMod2008, BrzZusMod2008-ToN} brings some attention to multi-hop (routing) definitions for LoP.
Brzezinski \ea~\cite{BrzZusMod2008-ToN} investigate scheduling on arbitrary graphs by decomposing, or pre-partitioning, the graph topology into multiple `orthogonal' trees and then applying known LoP results about GMS optimality on trees.
Both Joo \ea~\cite{JooLinShr2009-ToAC} and Kang \ea~\cite{KanJarYin2015} also treat the case of multi-hop traffic and LoP conditions.

\subsection{Motivation \& Contributions}\label{sec:motivation}

Much of the work reviewed above focuses on the issue of identifying the performance of GMS via the LoP factor for a given graph or select classes of graphs.
However, aside from the worst-case LoP analysis in geometric-unit-disk graphs by Joo \ea~\cite{JooLinShr2009-ToN} we are not aware of any work on establishing statistics and trends on the LoP factor \(\sigma\) in networks modeled as random graphs.
We note that the topology and structure of random graphs families, such as \Erdos-\Renyi (ER) and random geometric (RG) graphs, are tightly coupled with the density of edges present in the graph.
Our paper seeks to fill this void by rigorously establishing relationships between network edge densities and the resulting LoP factor in networks modeled as random graphs.
When viewed within the context of Joo \ea~\cite{JooLinShr2009-ToN}, statistics on the LoP factor \(\sigma\) are equivalent to statistics on \(\gamma^*\), the relative size of GMS's stability (or capacity) region.
We then place LoP within a larger context of commonly studied properties in both random graph families by comparison with the likelihood of connectivity properties.

Our paper and contributions are organized as follows.
In \secref{model}, we introduce our network model and provides essential definitions of threshold functions and the graph properties of interest.
In \secref{er-graphs}, we examine ER graphs due to their analytical tractability and gain insight into the behavior of the LoP factor relative to the chosen edge probability function.
We establish a regular threshold function based on the forbidden subgraph characterization of LoP \cite{BirChuRie2012} that dictates the likelihood that a graph satisfies LoP \(\sigma=1\) conditions (\thmref{er-plop-threshold}) and carry this analysis into bounds on the expected LoP factor (\thmref{er-sigma-exp-bounds}).
In \secref{rg-graphs}, we extend our analysis to the case of RG graphs due to their natural connection to wireless network models.
While the spatial dependence between edges in RG graphs complicates analysis, we are able to establish an upper bound for LoP threshold function (\prpref{rg-plop-upperbound} and \corref{rg-plop-0-statement}) as well as similar bounds on the expected LoP factor (\thmref{rg-sigma-exp-bounds}).
In both ER and RG sections, the LoP threshold functions are shown to produce a mutual exclusion between LoP and notions of connectedness (giant components and traditional connectivity) as the size of the network grows, for a large class of edge probability/radius functions (\thmref{er-exclusion-plop-pgiant} and \corref{er-exclusion-plop-pconn} for ER graphs; \thmref{rg-exclusion-plop-pgiant} and \corref{rg-exclusion-plop-pconn} for RG graphs).
).
In \secref{algorithms}, we comment on aspects of our numerical results, particularly on algorithm implementation to detect necessary or sufficient conditions for LoP in \iid realizations of ER and RG graphs.
In \secref{results}, we compare the analytical mutual exclusion of LoP and giant components with that of numerical results for finite network sizes and find that convergence to this exclusion between properties is rather quick as the network grows in size.
In \secref{conclusions}, we conclude our work and touch upon ideas for future investigation.
Finally, for clarity, long proofs are presented in the Appendix.

\section{Model \& Definitions}\label{sec:model}

Let \(\GGn\) be the set of all \(2^{\binom{n}{2}}\) simple graphs on \(n\) nodes.
A common variant of an \Erdos{}-\Renyi{} (ER) graph is constructed from \(n\) nodes where undirected edges between pairs of nodes are added using \iid{} Bernoulli trials with \emph{edge probability} \(p \in [0,1]\).
For each choice of \(p\), let \(\GGnp\) denote the finite probability space formed over \(\GGn\).

We will also consider a common variant of a random geometric (RG) graph, in which \(n\) node positions are modeled by a Binomial Point Process (BPP) within a unit square \([-1/2,1/2]^2 \subset \Rbb^2\).
Undirected edges between pairs of nodes are added iff the Euclidean distance between the two nodes is less than a given, fixed distance \(r \in [0,\infty)\).
For each choice of \(r\), let \(\GGnr\) denote the finite probability space formed over \(\GGn\).
Note that the particular RG model we have chosen is equivalent to a Poisson Point Process (PPP) conditioned on having \(n\) nodes within the unit square, producing an `equivalent' intensity \(\lambda = n\).

Interference in a graph \(\Gn \in \GGn\) is captured as a pairwise function between its edges.
Specifically, we adopt the primary (one-hop) interference model, under which adjacent edges (sharing a common node) interfere with one another.
Under this assumption, we can employ the forbidden subgraph characterization of LoP conditions found in \cite{BirChuRie2012}.

Let \(\Prop\) refer to both \emph{i)} a specific property or condition of a graph \(\Gn\), as well as \emph{ii)} the subset of graphs of \(\GGn\) for which the property holds, as described by \defref{graph-property}.
\begin{definition}[Graph Property \cite{Die2005}]\label{def:graph-property}
  A \emph{graph property} \(\Prop\) is a subset of \(\GGn\) that is closed under isomorphism (\(\sim_{\textup{iso}}\)): \ie, \(G \in \Prop, H \in \GGn, G \sim_{\textup{iso}} H \Rightarrow H \in \Prop\).
\end{definition}
\begin{definition}[Monotone Graph Property \cite{Bol2001}]\label{def:monotonic-property}
  Graph property \(\Prop\) is \emph{monotone increasing} if \(G \in \Prop, H \supset G \Rightarrow H \in \Prop\).
  Correspondingly, graph property \(\Prop\) is \emph{monotone decreasing} if \(G \in \Prop, H \subset G \Rightarrow H \in \Prop\).
\end{definition}

Let \(\Prob{\Gnp \in \Prop}\) denote the probability that a random graph \(\Gnp\) generated according to \(\GGnp\) satisfies graph property \(\Prop\).
For a monotone (increasing or decreasing) graph property, \(\Prop\), increasing the edge probability \(p \in [0,1]\) will cause a corresponding transition of \(\Prob{\Gnp \in \Prop}\) between \(0\) and \(1\).
Similarly, \(\Prob{\Gnr \in \Prop}\) (analogously defined using \(\Gnr\) and \(\GGnr\)) for a monotone graph property will also experience a transition as the edge distance \(r \in [0,\infty)\) increases.
In this case, it is of interest to study the behavior of the limiting probability \(\limninfty \Prob{\Gnpn \in \Prop}\) and \(\limninfty \Prob{\Gnrn \in \Prop}\) in response to the choice of \(\pn\) and \(\rn\), respectively.
We will use \(\Prob{\Prop}\) as a short form for \(\Prob{\Gnpn \in \Prop}\) or \(\Prob{\Gnrn \in \Prop}\) and use a general edge function \(\en\) as a stand in for either \(\pn\) or \(\rn^2\).
Note, thresholds of RG graphs on \(\Rbb^2\) are more easily expressed as the square of the edge distance \(\rn^2\) as opposed to \(\rn\).
A threshold function \(\esn\) for graph property \(\Prop\), when it exists, helps determine the limiting behavior of \(\Prob{\Prop}\) for choices of edge function \(\en\) relative to \(\esn\).
As in \cite{JanLucRuc2000}, we use the phrase `\(\Prop\) holds asymptotically almost surely (\aas)' to mean \(\limninfty \Prob{\Prop} = 1\) and the phrase `\(\Prop\) holds asymptotically almost never (\aan)' to mean \(\limninfty \Prob{\Prop} = 0\).
The asymptotic equivalence of two functions is denoted \(f(n)\!\sim\!g(n)\), that is \(\limninfty f(n) / g(n) = 1\).
We use the phrase `asymptotically positive' to describe \(f(n)\) if \(\exists n_0 : f(n) > 0, \forall n > n_0\).
Finally, let \(\Phi(x)\) be the \cdf{} of a standard normal \rv{}, and let \(n^{\underline{k}} = n!/(n-k)!\) denote the falling factorial.

\subsection{Threshold Functions}

First, we restate threshold function definitions in \cite{ErdRen1960} for a graph property \(\Prop\) using edge function \(\en\), threshold function \(\esn\), and the asymptotic notation of \cite{CorLeiRiv2009}.
\begin{definition}[Threshold Function]\label{def:threshold-fn}
  \(\esn\) is a \emph{threshold function} for monotonically increasing graph property \(\Prop\) if:
  \begin{equation}\label{eq:threshold-fn}
    \limninfty \Prob{\Prop} =
    \begin{cases}
      0, & \en \in \oh{\esn} \\
      1, & \en \in \om{\esn}
    \end{cases}.
  \end{equation}
\end{definition}
\begin{definition}[Regular Threshold Function]\label{def:reg-threshold-fn}
  \(\esn\) can be called a \emph{regular threshold function} if there exists a distribution function \(F(x)\) for \(0 < x < \infty\) such that at any of \(F\)'s points of continuity, \(x\):
  \begin{equation}
    \en \sim x \esn \Rightarrow \limninfty \Prob{\Prop} = F(x).
  \end{equation}
  \(F(x)\) is known as the \emph{threshold distribution function} for graph property \(\Prop\).
\end{definition}

When satisfied, \defref{threshold-fn} covers the limiting behavior of \(\Prob{\Prop}\) for all \(\en\) that lie an order of magnitude away from threshold \(\esn\).
Conversely, any function \(\en \in \Th{\esn}\) is also a threshold function of graph property \(\Prop\).
\defref{threshold-fn} has also been called a weak, or coarse, threshold function \cite{McC2004, Die2005, JanLucRuc2000}.
When \defref{reg-threshold-fn} applies, we can control the limiting value of \(\Prob{\Prop}\) to the extent that \(F(x)\) allows.
This can be accomplished by choosing \(\en\) to be a multiplicative factor \(x\) of \(\esn\).

The two `statements' of a threshold function:
\begin{align*}
  \en \in \oh{\esn} \Rightarrow & \limninfty \Prob{\Prop} = 0 \\
  \en \in \om{\esn} \Rightarrow & \limninfty \Prob{\Prop} = 1,
\end{align*}
are commonly referred to as the \(0\)-statement and the \(1\)-statement, as they dictate when \(\Prop\) holds with limiting probability \(0\) or \(1\).
For a monotone decreasing property, the \(0\)- and \(1\)-statements are appropriately reversed.

\subsection{Sharp Threshold Functions}

Stronger variations of the weak threshold have been defined, called either sharp, strong, or very strong threshold functions \cite{McC2004, HanMak2007, JanLucRuc2000}.
We restate sharp threshold function definitions in \cite{ErdRen1960} using \(\en\), \(\esn\), and the asymptotic notation of \cite{CorLeiRiv2009}.
\begin{definition}[Sharp Threshold Function]\label{def:sharp-threshold-fn}
  A \((\esn,\an)\) pair is a \emph{sharp threshold function} for monotonically increasing graph property \(\Prop\) if \(\an \in o(\esn)\), \(\an\) is asymptotically positive, and:
  \begin{equation}
    \limninfty \Prob{\Prop} =
    \begin{cases}
      0, & \en \in \esn - \om{\an} \\
      1, & \en \in \esn + \om{\an}
    \end{cases}.
  \end{equation}
\end{definition}

When satisfied, \defref{sharp-threshold-fn} covers the limiting behavior of \(\Prob{\Prop}\) for all \(\en\) that lie an additive factor (greater than order \(\an\)) away from \(\esn\).
Conversely, any function \(\en \in \esn + \Oh{\an}\) is also a sharp threshold function of graph property \(\Prop\).
Also note: by itself, \(\esn\) is a regular threshold function, that is, \(\esn\) satisfies \defref{reg-threshold-fn} with `degenerate' distribution function \(F(x) = \mathbf{1}\{x>1\}\) \cite{ErdRen1960}.
When presented alone (without \(\an\)), \(\esn\) is still referred to as a sharp/strong threshold function \cite{JanLucRuc2000}, perhaps prompting \cite{HanMak2007} to propose the term `very strong' to denote a \((\esn,\an)\) pair.
\begin{definition}[Regular Sharp Threshold Function]\label{def:reg-sharp-threshold-fn}
  A sharp threshold function \((\esn,\an)\) is a \emph{regular sharp threshold function} if there exists a distribution function \(F(x)\) for \(-\infty < x < \infty\) such that for any of \(F\)'s points of continuity, \(x\):
  \begin{equation}\label{eq:sharp-form}
    \en \sim \esn + x\an \Rightarrow \limninfty \Prob{\Prop} = F(x).
  \end{equation}
  \(F(x)\) is known as the \emph{sharp-threshold distribution function} for graph property \(\Prop\).
\end{definition}

When \defref{reg-sharp-threshold-fn} applies, we may control the limiting value of \(\Prob{\Prop}\) to the extent that \(F(x)\) allows by choosing \(\en\) to be \(\esn\) plus a term asymptotically equivalent to \(x \an\).

\subsection{Graph Properties}

We are interested in several graph properties listed in \tabref{properties}.
We first list results from Birand \ea \cite{BirChuRie2012} establishing \emph{i)} a set of forbidden subgraphs that characterizes Local Pooling \(\Plop\) under primary interference constraints, and \emph{ii)} a simple upper bound on the number of edges permitting Local Pooling, \(\Pedge\).
We then establish some useful properties and bounds of \(\Plop\), namely separate sufficient and necessary properties for Local Pooling, \(\Plopl\) and \(\Plopu\).
Later, thresholds for these three properties \(\Pedge\), \(\Plopl\), and \(\Plopu\) will be compared with thresholds for two connectivity properties, \(\Pconn\) and \(\Pgiant\).
\begin{table}[!t]
  \renewcommand\arraystretch{1.1}
  \caption{Graph Properties}
  \label{tab:properties}
  \centering
  \begin{tabular}{|l|l|}
    \hline Symbol      & Property \\ \hline
    \(\Plop\)          & satisfies LoP (\thmref{plop}) \\
    \(\Pedge\)         & contains no more than \(2n\) edges \\
    \(\Plopl\)         & contains no cycles \\
    \(\Plopu\)         & contains no cycles of lengths \(\{k \geq 6, k \neq 7\}\) \\
    \(\Pconn\)         & is connected \\
    \(\Pgiant(\beta)\) & largest component has normalized size \(\geq \beta, \beta \in (0,1)\) \\
    \(\Pgiant\)        & \(\exists \beta > 0\): largest component has normalized size \(\geq \beta\) \\ \hline
  \end{tabular}
\end{table}
\begin{theorem}[Local Pooling \(\Plop\) {\cite[Thm.~3.1]{BirChuRie2012}}]\label{thm:plop}
  A graph \(\Gn \in \Plop\) if and only if it contains no subgraphs within the set \(\Fmc = \{C_k | k \geq 6, k \neq 7\} \cup \{D_{k}^{s,t} | k \geq 0; s,t \in \{5,7\}\}\), where \(C_k\) is a cycle of length \(k \geq 3\) and \(D_{k}^{s,t}\) is a union of cycles of lengths \(s\) and \(t\) joined by a \(k\)-edge path (a `dumbbell').
\end{theorem}
\begin{lemma}[\(\Pedge\) Necessary for \(\Plop\) {\cite[Lem.~3.6]{BirChuRie2012}}]\label{lem:pedge-necessary}
  \(\Pedge\) is a necessary condition for graph property \(\Plop\).
\end{lemma}
\begin{lemma}[\(\Plop\) Monotonicity]\label{lem:plop-monotonicity}
  \(\Plop\) is a monotone decreasing property.
\end{lemma}
\begin{IEEEproof}
  See \prfref{plop-monotonicity}.
\end{IEEEproof}

Since \(\Plop\) is a monotone property, we are assured of the existence of a threshold function (for both ER graphs \cite[Thm.~1.24]{FriKal1996} and RG graphs \cite[Thm.~1.1]{GoeRaiKri2005}).
While we establish a regular threshold for \(\Plop\) in ER graphs (\thmref{er-plop-threshold}), we note that a threshold function for \(\Plop\) in RG graphs is not currently known to us.
In the latter case, separate necessary and sufficient conditions bound the subset \(\Plop\) (\lemref{plop-prop-bounds}) as well as the probability \(\Prob{\Plop}\) (\lemref{plop-prob-bounds}).
These bounds will hold regardless of the random graph model (ER or RG) employed, and are used later in our numerical results (\secref{results}).
\begin{lemma}[Separate Sufficient and Necessary Conditions for \(\Plop\)]\label{lem:plop-prop-bounds}
  \(\Plopl\) and \(\Plopu\) are sufficient and necessary properties for \(\Plop\), respectively, producing nested subsets:
  \begin{equation}
    \Plopl \subseteq \Plop \subseteq \Plopu.
  \end{equation}
\end{lemma}
\begin{IEEEproof}
  See \prfref{plop-prop-bounds}.
\end{IEEEproof}
\begin{lemma}[Probability Bounds for \(\Plop\)]\label{lem:plop-prob-bounds}
  Under any choice of \(\pn\) (\(\rn\)) used to generate ER (RG) graphs on \(n\) nodes:
  \begin{equation}
    \Prob{\Plopl} \leq \Prob{\Plop} \leq \Prob{\Plopu}, \quad \forall n \in \Zbb^+.
  \end{equation}
\end{lemma}
\begin{IEEEproof}
  See \prfref{plop-prob-bounds}.
\end{IEEEproof}

We will also look to establish statistics on the LoP factor, \(\sigma \in [0,1]\), for specific random graph families.
In this regard, \lemref{sigma-bounds} and \lemref{sigma-C6k} will prove helpful.
\begin{lemma}[\(\sigma\)-LoP Bounds \cite{LinShr2006, JooLinShr2009-ToAC}]\label{lem:sigma-bounds}
  For an arbitrary graph \(G\), its LoP factor \(\sigma(G)\) adheres to the following bounds:
  \begin{equation}
    \frac{1}{2} \leq \sigma(G) \leq \sigma(H), \quad \forall H \subseteq G.
  \end{equation}
\end{lemma}
\begin{IEEEproof}
  The lower bound of \(1/2\) is immediate from \cite{LinShr2006}.
  The upper bound follows from \cite[Def.~2.5]{BirChuRie2012}, a reformulation of \cite[Def.~6]{JooLinShr2009-ToAC}.
\end{IEEEproof}

\begin{lemma}[\(\sigma\)-LoP of \(C_{6k}\) \cite{BirChuRie2012}]\label{lem:sigma-C6k}
  \(\sigma(C_{6k}) = 2/3, \forall k \in \Nbb_+\).
\end{lemma}
\begin{IEEEproof}
  Under primary interference, the interference graph of \(G\) is its line graph.
  The line graph of any cycle \(C_k\) is itself.
  The result follows immediately by a specialization of \cite[Lem.~5.1]{BirChuRie2012} with \(6k\) in place of \(n\).
\end{IEEEproof}

\section{ER Graphs}\label{sec:er-graphs}

In this section, we examine several properties of interest for ER graphs.
We first provide a regular sharp threshold function for \(\Pedge\), a necessary property for \(\Plop\).
We also find that a regular threshold and distribution function can be directly established for property \(\Plop\) by considering the presence of forbidden subgraphs in \(\Fmc\).
We extend this argument to bound the support of the LoP factor \(\sigma(\Gnpn)\) as well as its expectation.
Known threshold functions for connectivity and giant components are re-stated for comparison with that of \(\Plop\).
We show that the threshold function for \(\Plop\) is incompatible with the known regular threshold function for \(\Pgiant(\beta)\) --- that is, choosing \(\pn\) so that \(\Pgiant(\beta)\) holds \aas{} implies that \(\Plop\) holds \aan.
It then follows that the stricter notion of connectivity is also incompatible with \(\Plop\).

\subsection{Local Pooling}

If we want to keep the expected number of edges in \(\Gnpn\) to be exactly \(2n\), we should set \(\pn = 4/(n-1)\).
This naturally suggests a threshold function of \(\psn = 1/n\).
This is indeed a threshold function for \(\Pedge\) (as are \(\psn = 4/(n-1)\) and \(\psn = 4/n\)).
While not particularly novel, we include \prpref{er-pedge-threshold} as we have not come across a citation for the result.

\begin{proposition}[Regular Sharp Threshold for \(\Pedge\) in \(\Gnpn\)]\label{prp:er-pedge-threshold}
  The pair \((\psn = 4 / n, \an = 2\sqrt{2n} / n^2)\) is a regular sharp threshold function for graph property \(\Pedge\) with distribution function \(F(x) = \Phi(-x)\) (flipped Normal).
\end{proposition}
\begin{IEEEproof}
  See \prfref{er-pedge-threshold}.
\end{IEEEproof}

Note, the condition \(\Pedge\) is not sufficient for \(\Plop\) and only provides an upper bound on a threshold function for \(\Plop\).
We improve upon this by considering established thresholds for the presence of individual forbidden subgraphs (such as cycles and dumbbells) in \(\Gnpn\).
Note, the threshold for the existence of edge-induced subgraphs in ER graphs is related to the maximum density of edges to vertices of the subgraph \cite{JanLucRuc2000}.
Cycles of a given length, being less `dense', will tend to occur at a lower threshold \(\pn \sim 1/n\) than dumbbells.
By focusing on just the set of forbidden cycles, we find that these individual thresholds combine to form a `semi-sharp' regular threshold function for \(\Plop\), similar in form to the threshold for all cycles \cite[Thm.~5b]{ErdRen1960}.
This is formalized by \thmref{er-plop-threshold}.

\begin{theorem}[Regular Threshold for \(\Plop\) in \(\Gnpn\)]\label{thm:er-plop-threshold}
  \(\psn = 1/n\) is a regular threshold function for graph property \(\Plop\), with distribution function:
  \begin{equation}
    F(x) =
    \begin{cases}
      \sqrt{1-x}\exp\left(\sum_{k \in \Kmc}\frac{x^k}{2k} \right), & x < 1 \\
      0, & x \geq 1
    \end{cases}
  \end{equation}
  where \(\Kmc = \{1,2,3,4,5,7\}\).
\end{theorem}
\begin{IEEEproof}
  See \prfref{er-plop-threshold}.
\end{IEEEproof}

\thmref{er-plop-threshold} provides the limiting behavior of \(\Prob{\Plop}\) when \(\pn\) is chosen relative to \(1/n\).
In the case that \(\pn\) is asymptotically larger than \(1/n\), we have that \(\Plop\) is satisfied \aan{}.
However, in order to guarantee that \(\Plop\) is satisfied \aas{}, \(\pn\) must be chosen \(\oh{1/n}\).
Thus, we have established how to choose \(\pn\) in order to asymptotically satisfy \(\Plop\) with probability between \(0\) and \(1\).
Correspondingly, \thmref{er-plop-threshold} can be weakened to provide a threshold function for property \(\Plop\).
\begin{corollary}[Threshold Function for \(\Plop\) in \(\Gnpn\)]\label{cor:er-plop-threshold-weak}
  \(\psn = 1/n\) is a threshold function for \(\Plop\).
\end{corollary}
\begin{IEEEproof}
  See \prfref{er-plop-threshold-weak}.
\end{IEEEproof}

In dense networks above the threshold \(\psn = 1/n\), we find that the support for the LoP factor is bounded between \(1/2\) and \(2/3\):
\begin{proposition}[\(\sigma\)-LoP Bounds in \(\Gnpn\)]\label{prp:er-sigma-prob-bounds}
  When \(\pn \sim c/n, c>1\), the limiting behavior of the LoP factor \(\sigma\) may be bounded as follows:
  \begin{equation}
    \limninfty \Prob{1/2 \leq \sigma(\Gnpn) \leq 2/3} = 1.
  \end{equation}
\end{proposition}
\begin{IEEEproof}
  See \prfref{er-sigma-prob-bounds}.
\end{IEEEproof}

\begin{theorem}[\(\Expect{\sigma}\) Bounds in \(\Gnpn\)]\label{thm:er-sigma-exp-bounds}
  Let \(\pn \sim c/n\).
  The limiting behavior of \(\Expect{\sigma(\Gnpn)}\) may be bounded by.
  \begin{equation}
    \frac{1}{2}\left(1 + F_l(c)\right) \leq \limninfty \Expect{\sigma(\Gnpn)} \leq \frac{1}{3}\left(2 + F_u(c)\right)
  \end{equation}
  where:
  \begin{equation}
    F_l(x) = \begin{cases}
      \sqrt{1-x} \exp\left(\sum_{k\in\Kmc} \frac{x^k}{2k}\right), & x < 1 \\
      0, & x \geq 1
    \end{cases},
  \end{equation}
  \begin{equation}
    F_u(x) = \begin{cases}
      (1-x^6)^{1/12}, & x < 1 \\
      0, & x \geq 1
    \end{cases}
  \end{equation}
  with \(\Kmc = \{1,2,3,4,5,7\}\).
\end{theorem}
\begin{IEEEproof}
  See \prfref{er-sigma-exp-bounds}.
\end{IEEEproof}
Note, when \(\pn \sim c/n\) and \(c>1\), bounds on the expected value of \(\sigma(G)\) are a primarily a function of the restricted support provided by \prpref{er-sigma-prob-bounds}.
A visualization of the bounds is provided in \figref{sigma-bounds}, which prove to be quite tight for \(c<1\).

\subsection{Connectivity and Giant Components}

Previously established results provide a sharp threshold function for connectivity in ER graphs:
\begin{lemma}[Regular Sharp Threshold for \(\Pconn\) in \(\Gnpn\) \cite{ErdRen1959,Bol2001}]\label{lem:er-pconn-threshold}
  The pair \((\psn = \log(n)/n, \an = 1/n)\) is a regular sharp threshold function for graph property \(\Pconn\) with distribution function \(F(x) = \exp(-\exp(-x))\) (Gumbel).
\end{lemma}

We can also loosen our restriction that \(G\) be connected and look at threshold functions for the formation of giant components in random graphs.
A giant component exists if the largest connected components contains a positive fraction of the vertices of \(G\) as \(n \rightarrow \infty\).
Janson \ea provide a relevant threshold function \(\psn = c(\beta)/n\) for the existence of a giant component with normalized size \(\beta \in (0,1)\) \cite[Thm.~5.4]{JanLucRuc2000}.
We find that the same threshold function easily applies to the existence of a giant component of size at least \(\beta\).
\begin{corollary}[Regular Threshold for \(\Pgiant(\beta)\) in \(\Gnpn\) \cite{JanLucRuc2000}]\label{cor:er-pgiant-threshold}
  Let \(\beta^* \in (0,1)\), then \(\psn = c(\beta^*)/n\) is a regular threshold function for graph property \(\Pgiant(\beta^*)\), with distribution function \(F(x) = \mathbf{1}\{x > 1\}\), where:
  \begin{equation}\label{eq:c-beta}
    c(\beta) = \frac{1}{\beta}\ln\left(\frac{1}{1-\beta}\right).
  \end{equation}
\end{corollary}
\begin{IEEEproof}
  See \prfref{er-pgiant-threshold}.
\end{IEEEproof}

Given the facts that that \emph{i)} \(\Plop\) and \(\Pgiant\) are monotone decreasing and increasing, resp., and \emph{ii)} their respective threshold functions do not `overlap' (recall that \(c(\beta) > 1\)), we present a statement of mutual exclusion between the two properties:
\begin{theorem}[Mutual Exclusion of \(\Plop\) and \(\Pgiant(\beta)\) in \(\Gnpn\)]\label{thm:er-exclusion-plop-pgiant}
  In ER graphs with edge probability function \(\pn\) and desired giant component size \(\beta \in (0,1)\):
  \begin{equation}
    \limninfty \frac{\pn}{1/n} \geq 0 \Rightarrow \limninfty \Prob{\Plop \cap \Pgiant(\beta)} = 0.
  \end{equation}
\end{theorem}
\begin{IEEEproof}
  See \prfref{er-exclusion-plop-pgiant}.
\end{IEEEproof}

Note that the threshold for connectivity has a higher order than that of giant components (\(\log(n)/n\) vs. \(c(\beta)/n\)), thus, we expect (and find) that properties \(\Plop\) and \(\Pconn\) exhibit an identical mutual exclusion:
\begin{corollary}[Mutual Exclusion of \(\Plop\) and \(\Pconn\) in \(\Gnpn\)]\label{cor:er-exclusion-plop-pconn}
  In ER graphs with edge probability function \(\pn\):
  \begin{equation}
    \limninfty \frac{\pn}{1/n} \geq 0 \Rightarrow \limninfty \Prob{\Plop \cap \Pconn} = 0.
  \end{equation}
\end{corollary}
\begin{IEEEproof}
  \(\Pgiant(\beta)\) is necessary for connectivity \(\Pconn\), thus: \(\Prob{\Plop \cap \Pconn} \leq \Prob{\Plop \cap \Pgiant(\beta)}\).
  Mutual exclusion between \(\Plop\) and \(\Pconn\) follows immediately from \thmref{er-exclusion-plop-pgiant}.
\end{IEEEproof}

\begin{figure}[!t]
  \centering
  \includegraphics[width=0.8\columnwidth]{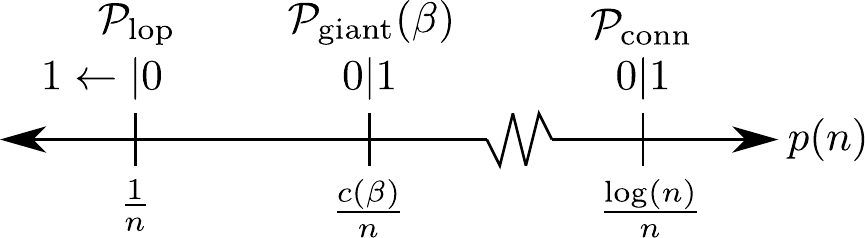}
  \caption{%
    The limiting behavior of the graph properties in \tabref{properties} along the design space of functions \(\pn\) chosen relative to established threshold functions for ER graphs.
    Listed from top to bottom are \emph{i)} the graph properties, \emph{ii)} their limiting probabilities relative to an established threshold function, \emph{iii)} the established threshold function.
  }
  \label{fig:er-prop-order}
\end{figure}

We note that the set of \(\pn\) covered by \thmref{er-exclusion-plop-pgiant} and \corref{er-exclusion-plop-pconn} is a rather large class, covering all functions that can be placed into an asymptotic relationship with \(1/n\).
This includes \(\oh{1/n}\) and \(\om{1/n}\), but leaves out certain functions that contain periodic components (\eg, \((\sin(n)+1) / n\)).
We note that these `sinusoidal' functions may oscillate across the threshold \(1/n\) for certain graph properties of interest and do not make sense to employ when attempting to satisfy monotone properties in ER graphs.
We also note that a more elegant, larger characterization of the set of \(\pn\) that satisfy this mutual exclusion may exist (particularly for \(\Pconn\), whose threshold lies at a higher order than that of \(\Plop\)).
Refer to \figref{er-prop-order} for a visual comparison of the limiting behavior of the properties in \tabref{properties} in ER graphs.

\section{RG Graphs}\label{sec:rg-graphs}

In this section, we examine several properties of interest for RG graphs.
We first provide a regular sharp threshold function for \(\Pedge\), a necessary property and threshold upper bound for \(\Plop\).
We obtain a tighter threshold upper bound for \(\Plop\) by considering the presence of forbidden subgraphs in \(\Fmc\).
This upper bound is sufficient to prove the threshold for \(\Plop\) is incompatible with known regular threshold function \(\rsn^2 = \log(n)/(\pi n)\) for \(\Pconn\) --- that is, choosing \(\rn^2\) so that \(\Plop\) holds \aas{} implies that \(\Pconn\) holds \aan.
Further, relaxing our desire for connectivity from \(\Pconn\) to \(\Pgiant\) lowers the regular threshold function from \(\log(n)/(\pi n)\) to \(\lambda_c/n\) with \(\lambda_c \in (0,\infty)\).
However, we find that this is insufficient to prevent the incompatibility of \(\Plop\) with \(\Pgiant\).

\subsection{Local Pooling}

\begin{proposition}[Regular Sharp Threshold for \(\Pedge\) in \(\Gnrn\)]\label{prp:rg-pedge-threshold}
  The pair \((\rsn^2 = 4 / (\pi n), \an = 2\sqrt{2n} / (\pi n^2))\) is a regular sharp threshold function for graph property \(\Pedge\) with sharp-threshold distribution function \(F(x) = \Phi(-x)\) (flipped Normal).
\end{proposition}
\begin{IEEEproof}
  See \prfref{rg-pedge-threshold}.
\end{IEEEproof}

\begin{remark}
  The leading term of the threshold in \prpref{rg-pedge-threshold} was motivated by solving an expression for the expected number of edges in \(\Gnrn\) for \(\rn^2\).
  The second term of the threshold is specifically chosen such that all multiplicative factors other than \(-x\) cancel out from scaling \((a)\) and subsequent standardization \((d)\) in the proof.
\end{remark}

\begin{proposition}[Upper Bound for \(\Plop\) in \(\Gnrn\)]\label{prp:rg-plop-upperbound}
  When \(\rsn^2 \sim c/n^{6/5}\), an upper bound for LoP may be expressed:
  \begin{equation}
    \limsupninfty \Prob{\Plop} \leq \exp\left(\frac{-(\pi c/4)^5}{6!}\right)
  \end{equation}
\end{proposition}
\begin{IEEEproof}
  See \prfref{rg-plop-upperbound}.
\end{IEEEproof}

\begin{remark}
  Unlike the case of ER graphs where cycles of all orders began appearing at the same threshold \(\pn \sim 1/n\), the RG thresholds of forbidden subgraphs in \(\Fmc\) are more spread out (order \(k\) vertex-induced subgraphs yielding an order \(k\) edge-induced forbidden subgraph begin to appear at \(\rn^2 \sim n^{-k/(k-1)}\)).
  For \prpref{rg-plop-upperbound}, we wished to find the tightest upper bound for \(\Plop\) that was amenable to asymptotic analysis.
  Thus, we first restricted our attention to the lowest order vertex-induced subgraphs (subgraphs of order \(6\)).
  Second, we noted that evaluating \(\mu_{\Gamma_6}\) for all feasible, order \(6\) graphs that contain the forbidden edge-induced \(C_6\) appears to be neither analytically tractable nor computationally viable, so we apply a second upper bound by focusing on a specific vertex-induced subgraph, the complete graph \(K_6\), and derive an easy upper bound for \(\mu_{K_6}\).
\end{remark}

The upper bound in \prpref{rg-plop-upperbound} yields a \(0\)-statement:
\begin{corollary}[\(0\)-statement for \(\Plop\) in \(\Gnrn\)]\label{cor:rg-plop-0-statement}
  When \(\rn^2 \in \om{1/n^{6/5}}\), \(\limninfty \Prob{\Plop} = 0\).
\end{corollary}
\begin{IEEEproof}
  This follows immediately from \prpref{rg-plop-upperbound}.
\end{IEEEproof}

Due to fact that all forbidden subgraphs contain at least \(6\) or more vertices, it does not seem likely that a corresponding \(1\)-statement would hold at a lower threshold than \(\rsn^2 \sim 1/n^{6/5}\).
Thus, we are led to make the following conjecture:
\begin{conjecture}[Threshold for \(\Plop\) in \(\Gnrn\)]\label{cnj:rg-plop-threshold}
  \(\rsn^2 = 1/n^{6/5}\) is a threshold function for graph property \(\Plop\).
\end{conjecture}
The difficulty in proving this conjecture lies in establishing a sufficient condition whose probability lower bounds \(\Prob{\Plop}\) while maintaining enough tractability to take its limit as \(n\rightarrow\infty\).
In terms of applying the same proof strategy as used for ER graphs, we note that the FKG inequality does not appear readily applicable.
We note that none of the results presented in this paper depends on this conjecture.

Similar to the case for ER graphs, the asymptotic support for the LoP factor for dense RG networks (above the threshold \(\rsn^2 = 1/n^{6/5}\)) also lies between \(1/2\) and \(2/3\):
\begin{proposition}[\(\sigma\)-LoP Bounds in \(\Gnrn\)]\label{prp:rg-sigma-prob-bounds}
  When \(\rn^2 \in \om{1/n^{6/5}}\), the limiting behavior of the LoP factor \(\sigma\) may be bounded as follows:
  \begin{equation}
    \limninfty \Prob{1/2 \leq \sigma(\Gnrn) \leq 2/3} = 1.
  \end{equation}
\end{proposition}
\begin{IEEEproof}
  See proof in \prfref{rg-sigma-prob-bounds}.
\end{IEEEproof}
\begin{theorem}[\(\Expect{\sigma}\) Bounds in \(\Gnrn\)]\label{thm:rg-sigma-exp-bounds}
  Let \(\rn^2 \sim c/n^{6/5}\).
  We may bound the limiting behavior of \(\Expect{\sigma(\Gnrn)}\) as a function of \(x\).
  \begin{equation}
    \frac{1}{2} \leq \limninfty \Expect{\sigma(\Gnrn)} \leq \frac{1}{3}\left(2 + \exp\left(-\frac{(\pi c / 4)^{5}}{6!}\right)\right).
  \end{equation}
\end{theorem}
\begin{IEEEproof}
  See proof in \prfref{rg-sigma-exp-bounds}.
\end{IEEEproof}

\begin{figure}[!t]
  \centering%
  \includegraphics[width=0.5\columnwidth]{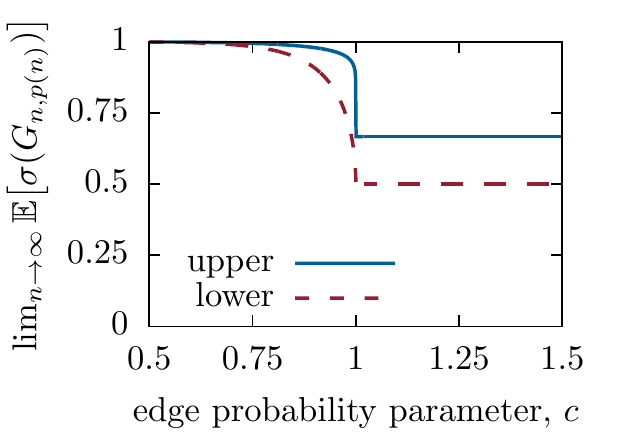}%
  \includegraphics[width=0.5\columnwidth]{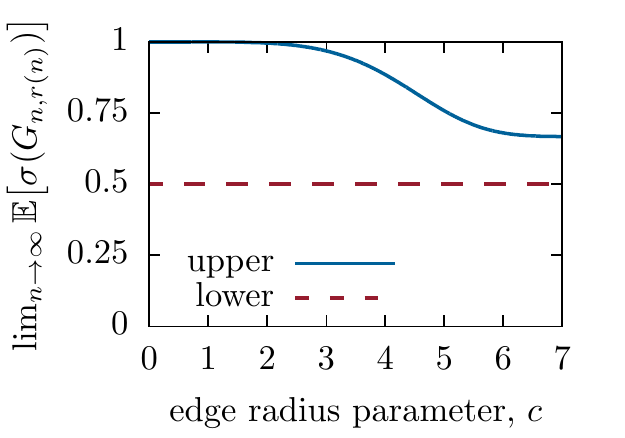}%
  \caption{%
    Limiting bounds (\(n\rightarrow\infty\)) on \(\Expect{\sigma(G)}\) in ER graphs when \(\pn \sim c/n\) (\textbf{left}), and RG graphs when \(\rn^2 \sim c/n^{6/5}\) (\textbf{right}).
  }
  \label{fig:sigma-bounds}
\end{figure}

In \figref{sigma-bounds}, we present a visual comparison of the limiting behavior of \(\Expect{\sigma}\) for both ER and RG graphs.
These bounds are provided by \thmref{er-sigma-exp-bounds} and \thmref{rg-sigma-exp-bounds}, respectively.
We note that the tighter bounds for ER graphs is afforded by the coinciding cycle subgraph thresholds at \(\pn \sim 1/n\).

\subsection{Connectivity and Giant Components}

Previously established results provide a regular sharp threshold function for connectivity and a regular threshold for giant components in RG graphs:
\begin{lemma}[Regular Sharp Threshold for \(\Pconn\) in \(\Gnrn\) \cite{Pen2003}]\label{lem:rg-pconn-threshold}
  The pair \((\rsn^2 = \log(n)/(\pi n), \an = 1/(\pi n))\) is a regular sharp threshold function for graph property \(\Pconn\) with sharp-threshold distribution function \(F(x) = \erm^{-\erm^{-x}}\) (Gumbel).
\end{lemma}
\begin{lemma}[Regular Threshold for \(\Pgiant\) in \(\Gnrn\) \cite{Pen2003}]\label{lem:rg-pgiant-threshold}
  \(\rsn^2 = \lambda_c/n\) is a regular threshold function for graph property \(\Pgiant\) with threshold distribution function \(F(x) = \mathbf{1}\{x > 1\}\), where \(\lambda_c \in (0,\infty)\) is the critical percolation threshold.
\end{lemma}

Given the facts that that \emph{i)} \(\Plop\) and \(\Pgiant\) are monotone decreasing and increasing properties respectively, and \emph{ii)} their respective \(0\)-statements `overlap', we present a statement of mutual exclusion between the two properties:
\begin{theorem}[Mutual Exclusion of \(\Plop\) and \(\Pgiant\) in \(\Gnrn\)]\label{thm:rg-exclusion-plop-pgiant}
  In RG graphs with edge radius function \(\rn\):
  \begin{equation}
    \limninfty \frac{\rn^2}{1/n} \geq 0 \Rightarrow \limninfty \Prob{\Plop \cap \Pgiant} = 0.
  \end{equation}
\end{theorem}
\begin{IEEEproof}
  See \prfref{rg-exclusion-plop-pgiant}.
\end{IEEEproof}

Again, in the case of RG graphs, the threshold for connectivity has a higher order than that of giant components (\(\log(n)/(\pi n)\) vs. \(\lambda_c / n\)), thus, we expect (and find) that properties \(\Plop\) and \(\Pconn\) exhibit an identical mutual exclusion:
\begin{corollary}[Mutual Exclusion of \(\Plop\) and \(\Pconn\) in \(\Gnrn\)]\label{cor:rg-exclusion-plop-pconn}
  In RG graphs with edge radius function \(\rn\):
  \begin{equation}
    \limninfty \frac{\rn^2}{1/n} \geq 0 \Rightarrow \limninfty \Prob{\Plop \cap \Pconn} = 0.
  \end{equation}
\end{corollary}
\begin{IEEEproof}
  \(\Pgiant\) is necessary for connectivity \(\Pconn\), thus \(\Prob{\Plop \cap \Pconn} \leq \Prob{\Plop \cap \Pgiant}\).
  Mutual exclusion between \(\Plop\) and \(\Pconn\) follows immediately from \thmref{rg-exclusion-plop-pgiant}.
\end{IEEEproof}

\begin{figure}[!t]
  \centering
  \includegraphics[width=0.8\columnwidth]{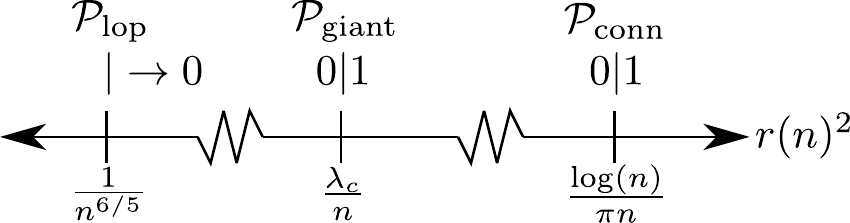}
  \caption{%
    The limiting behavior of the graph properties in \tabref{properties} along the design space of functions \(\rn^2\) chosen relative to established threshold functions for RG graphs.
    Listed from top to bottom are \emph{i)} the graph properties, \emph{ii)} their limiting probabilities relative to an established threshold function or \(0\)-statement in the case of \(\Plop\), \emph{iii)} the established threshold function.
  }
  \label{fig:rg-prop-order}
\end{figure}

In the case of RG graphs, we note that the threshold for \(\Plop\) must lie at a lower order than both that of \(\Pgiant\) and \(\Pconn\), whereas in ER graphs, \(\Plop\) and \(\Pgiant\) were both located at \(1/n\).
Refer to \figref{rg-prop-order} for a visual comparison of the limiting behavior of the properties in \tabref{properties} in RG graphs.

\section{Algorithms for Bounding \texorpdfstring{\(\Plop\)}{Plop}}\label{sec:algorithms}

Birand \ea~\cite{BirChuRie2012} outline an \(\Oh{n}\)-time exact algorithm checking whether or not a graph with \(n\) vertices satisfies \(\Plop\) under primary interference constraints.
At a high-level, the algorithm involves decomposition of the graph into bi-connected components and checking each component for certain characteristics; among these is a test for `long' cycles (in order to exclude forbidden cycle lengths).
Our analytical results suggest that the formation of cycles are the major factor prohibiting LoP in ER and RG random graphs, so we have implemented algorithms to check for necessary and sufficient conditions for LoP in random graphs (\(\Plopl\) and \(\Plopu\)).
Our simulations are performed in Matlab, where we make use of MatlabBGL \cite{Gle2009-PhD} for graph decomposition into connected components and depth-first-search.
These following algorithms and their supporting functions are listed in Listing \ref{alg:lop} and are centered around the detection of long cycles.

\textsc{HasCycleEq} accepts an input graph \(G\), a cycle-length \(k\), and a maximum number of iterations \(I\) and reports whether or not a cycle of length \(k\) exists within \(G\).
\textsc{HasCycleEq} relies directly upon a randomized algorithm, denoted \textsc{AYK}, proposed by Alon \ea~\cite[Thm.~2.2]{AloYusZwi1995}, which iteratively generates random, acyclic, directed subgraphs of \(G\) and tests for cycles via the subgraph's adjacency matrix.
If no cycles of length \(k\) are found after the \(I\)th iteration, we have \textsc{HasCycleEq} report that no length \(k\) cycles exist in \(G\), which may be a false negative.
As a result, \textsc{HasCycleEq} is suitable for use in upper-bounding the probability of the non-existence of forbidden cycles, namely in \textsc{PlopU}.

\textsc{HasCycleGeq} accepts an input graph \(G\), a minimum cycle-length \(K\), and a maximum number of iterations \(I\) and reports whether or not a cycle of length \(K\) or greater exists within \(G\).
In general, the decision problem formulation (also known as the long-cycle problem) is NP-hard, but polynomial for fixed-parameter \(k\).
We make use of a result by Gabow and Nie \cite[Thm.~4.1]{GabNie2008}; for \(K > 3\), depth-first-search \textsc{DFS} may be used to detect the existence of cycles of length longer than \(2K-4\) by examining the back-edges discovered by \textsc{DFS}.
Note, a \textsc{DFS} back-edge of length \(K-1\) implies the existence of a length \(K\) cycle.
Thus, if a `long' back-edge is found by \textsc{DFS}, we may report that such a cycle exists (line \(7\)).
In the event that \textsc{DFS} fails to detect long backedges, a long simple cycle (if it exists) will have length between \(K\) and \(2K-4\) \cite[Thm.~4.1]{GabNie2008}.
For each length \(k\) within this range, we call the randomized algorithm in \textsc{HasCycleEq}, thus \textsc{HasCycleGeq} may also report false negatives.
Alternately, when \(K = 3\), \textsc{HasCycleGeq} is an exact algorithm (lines \(8\)-\(10\) involving \textsc{HasCycleEq} are short-circuited) that checks for the existence of any cycle.
This is accomplished by running \textsc{DFS} and examining the resulting tree for back-edges of length \(2\) or longer.
In the event no such back-edges are found, we may conclude that graph \(G\) is cycle-free.

Finally, we discuss \textsc{PlopL} and \textsc{PlopU}.
\textsc{PlopL} checks for the existence of any cycles and calls \textsc{HasCycleGeq} directly.
For the reasons discussed above, \textsc{PlopL} is an exact (not randomized) algorithm and suitable for lower bounding the probability of satisfying LoP conditions.
\textsc{PlopU} checks for the existence of forbidden cycles.
For forbidden cycles of length \(6\), we call \textsc{HasCycleEq}, while for fobidden cycles of length \(8\) or longer, we call \textsc{HasCycleGeq}.
For this reason, the curves displayed for \(\Plopu\) in later figures are an upper bound for \(\Plopu\) (which can be improved by increasing the number of allowed iterations, \(I\)), but nevertheless yield valid upper bounds for \(\Plop\) and additionally demonstrate the mutual exclusivity between \(\Plop\) and \(\Pgiant\) in ER and RG graphs.

\begin{remark}\label{rem:tighter-lower-bound}
  One could obtain a tighter sufficient condition \(\Plopl\) (and thus a tighter lower bound) by restricting cycles of length \(k \geq 5\) instead of all cycles.
  We have not done so for the following reasons: \emph{i)} the use of \textsc{HasCycleGeq} with \(K=5\) will not produce an exact answer (but instead an upper bound on \(\Plopl\)), and \emph{ii)} we are more concerned and satisfied with characterizing an upper bound for \(\Plop\) and its interaction with connectivity requirements.
\end{remark}

\floatname{algorithm}{Listing}
\newcommand{\TDFS}{T_{\textsc{DFS}}}

\begin{algorithm}[t]
  \caption{Pseudo-code checking for \(\Plopl\) and \(\Plopu\)}
  \label{alg:lop}
  \begin{algorithmic}[5]
    \Function{HasCycleEq}{$G$,$k$,$I$}
      \State \Return \Call{AYK}{$G$,$k$,$I$}
    \EndFunction
    \Function{HasCycleGeq}{$G$,$K$,$I$}
    \State $\TDFS \gets$ \Call{DFS}{$G$}
      \If {$\textsc{LongestBackedge}(\TDFS) \geq (K-1)$}
        \State \Return \textsc{true}
      \Else
        \For {$k = K$ to $2K-4$}
        \If {\Call{HasCycleEq}{$G$,$k$,$I$}}
            \State \Return \textsc{true}
          \EndIf
        \EndFor
      \EndIf
      \State \Return \textsc{false}
    \EndFunction
    \Function{PlopL}{$G$,$I$}
      \State \Return \Call{HasCycleGeq}{$G$,$3$,$I$}
    \EndFunction
    \Function{PlopU}{$G$,$I$}
    \If {\Call{HasCycleGeq}{$G$,$8$,$I$}}
      \State \Return \textsc{true}
    \Else
      \State \Return \Call{HasCycleEq}{$G$,$6$,$I$}
    \EndIf
    \EndFunction
  \end{algorithmic}
\end{algorithm}

\section{Numerical Results}\label{sec:results}

The analytical results presented thus far are asymptotic (\(n\rightarrow\infty\)).
In this section, we compare the analytical mutual exclusion of LoP and giant components with that of numerical results for finite network sizes and find that this exclusion occurs rather quickly as the network grows in size.

\subsection{ER Graphs}

\begin{figure}[!t]
  \centering%
  \includegraphics[width=\columnwidth]{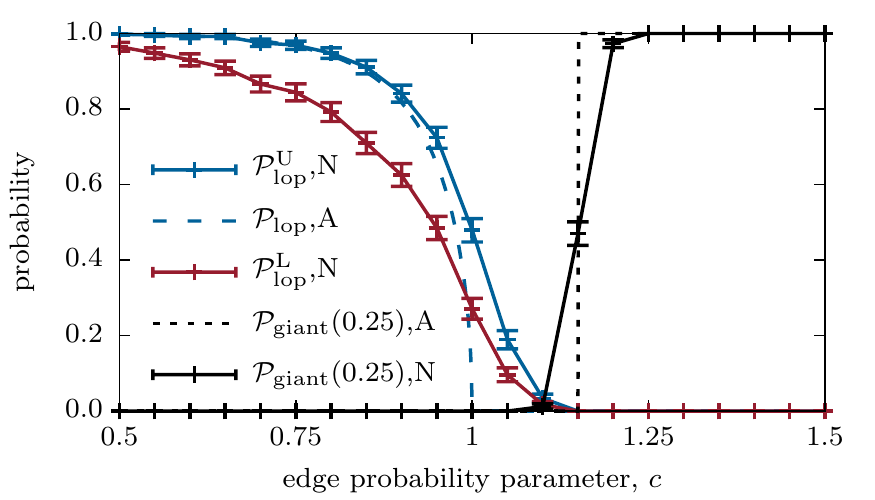}%
  \caption{%
    Probabilities of graph properties occurring in ER graphs are plotted as a function of \(c\) where the edge probability is chosen according to \(\pn = c/n\).
    Asymptotic (as \(n\rightarrow\infty\)), analytical (A) probabilities are plotted in dashed lines.
    Numerical (N) probabilities are plotted in solid lines with \(95\%\) confidence intervals generated from \(S = 10^3\) \iid graphs of size \(n = 10^4\).
    \textsc{PlopU} was configured to use a maximum of \(I = 10^4\) iterations.
  }
  \label{fig:er-prop-vs-c}
\end{figure}

In \figref{er-prop-vs-c}, we see that the numerical results generally match their analytical limits at \(n = 10^4\).
In particular, as \(n\rightarrow\infty\), the numerical curves associated with \(\Pgiant(\beta)\) become increasingly sigmoidal about \(c \approx 1.15\) when \(\beta\) is set to a rather conservative value of \(0.25\).
Also note that the effect of increasing the minimum required giant component size \(\beta\) serves to shift the associated curves in \figref{er-prop-vs-c} to the right, further negating any chance of both satisfying local pooling and having a giant component.
Regarding \(\Plop\), when \(c < 1\), we note that there is good agreement with the numerical upper bound and the gap with the lower bound is readily explained by \remref{tighter-lower-bound}.
When \(c > 1\), there are noticeable `tails' on the numerical bounds, and we are inclined to attribute the existence of the tails to the notion that graphs of a finite size \(n\) may only reliably capture the limiting behavior of small cycles, perhaps much smaller than \(n\).%
\footnote{By appropriately restricting conditions \(\Plopl\) and \(\Plopu\) to cycles of lengths less than finite \(K \approx 20\), the resulting threshold distribution functions more closely match the presented numerical results.}

\begin{figure}[!t]
  \centering%
  \includegraphics[width=\columnwidth]{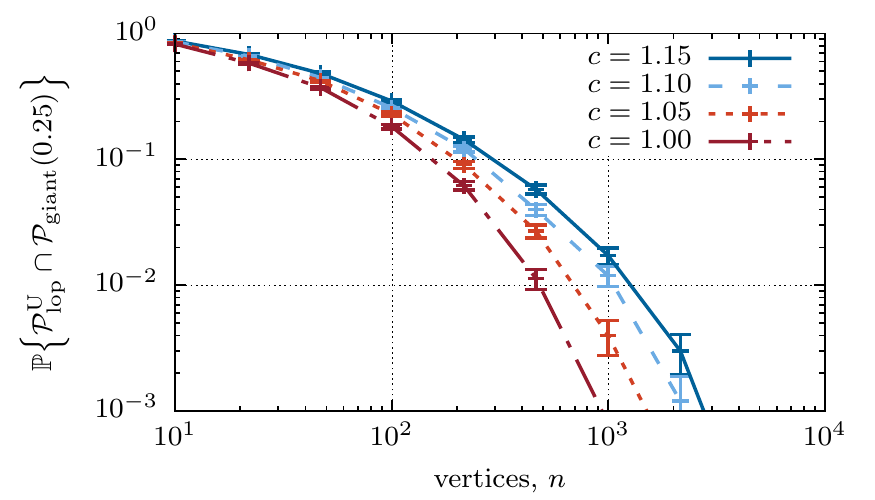}%
  \caption{%
    Numerical probability of satisfying both \(\Plopu\) and \(\Pgiant(0.25)\) in ER graphs plotted as a function of \(n\) where the edge probability is chosen according to \(\pn = c/n\).
    Numerical probabilities are computed with \(95\%\) confidence intervals generated from \(S = 10^4\) \iid graphs.
    \textsc{PlopU} was configured to use a maximum of \(I = 10^3\) iterations.
  }
  \label{fig:er-prop-vs-n}
\end{figure}

In \figref{er-prop-vs-n}, we focus on edge probability functions \(\pn = c/n\) with parameter \(1 \leq c \leq 1.15\), which falls between the asymptotic thresholds for \(\Plop\) and \(\Pgiant(0.25)\) (see \figref{er-prop-vs-c}).
For each edge probability function within this regime, we plot the probability that an ER graph satisfies both \(\Plopu\) and \(\Pgiant(0.25)\) as a function of the network size, \(n\).
We observe that the exclusion between \(\Plopu\) and \(\Pgiant\) develops rather rapidly.

\subsection{RG Graphs}

\begin{figure}[!t]
  \centering%
  \includegraphics[width=\columnwidth]{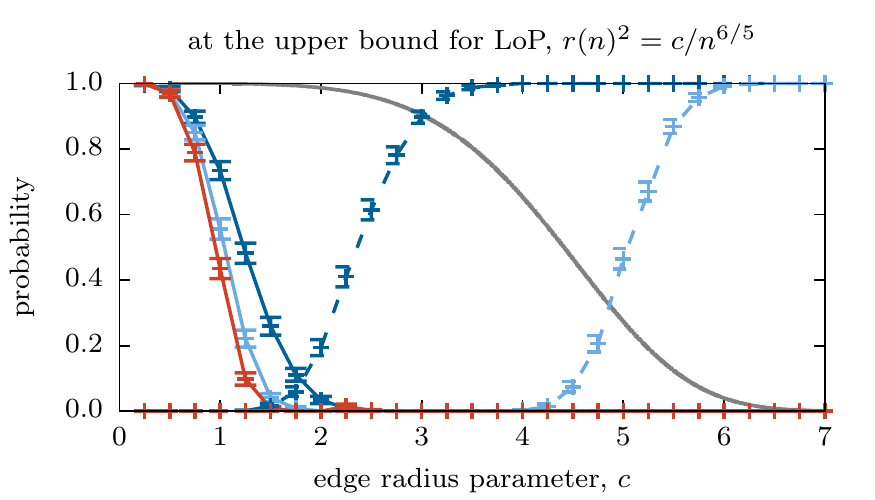}\\%
  \includegraphics[width=\columnwidth]{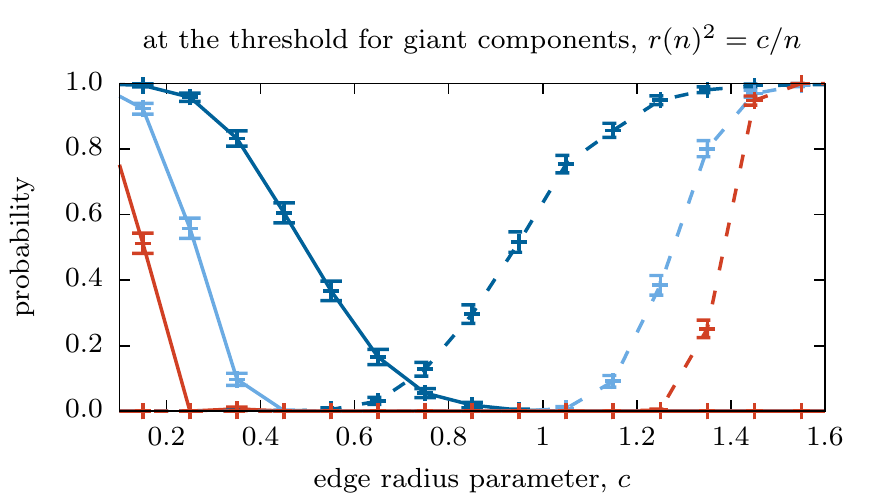}\\%
  \includegraphics[width=\columnwidth]{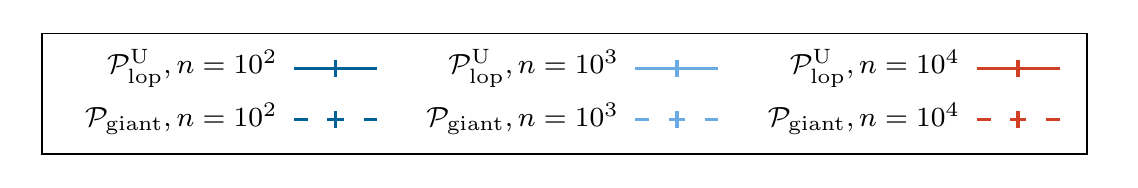}%
  \caption{%
    Probabilities of graph properties occurring in RG graphs are plotted as a function of \(c\) where the edge radius is chosen according to \(\rn^2 = c/n^{6/5}\) (\textbf{top}) and \(\rn^2 = c/n\) (\textbf{bottom}).
    Numerical (N) probabilities are plotted according to the legend with \(95\%\) confidence intervals generated from \(S = 10^3\) \iid graphs of sizes \(n = \{10^2,10^3,10^4\}\).
    \textsc{PlopU} was configured to use a maximum of \(I = 10^3\) iterations.
    An additional asymptotic upper bound for \(\Plop\) is plotted in solid grey for the top plot only.
  }
  \label{fig:rg-prop-vs-c}
\end{figure}

Unlike the case of ER graphs, we note that the RG graph bounds and thresholds for \(\Plop\) and \(\Pgiant(\beta)\) (respectively) must necessarily occur at edge radius functions of different orders of \(n\).
For this reason, we provide two subplots in \figref{rg-prop-vs-c} that are analogous to \figref{er-prop-vs-c} and separately consider edge radius functions \(\rn^2 = c/n^{6/5}\) and \(\rn^2 = c/n\).
Intuitively, for edge radius function \(\rn^2 = c/n^{6/5}\), we expect to see (and also observe) two phenomenon as the parameter \(n\) increases: the probability of \(\Plopu\) should show convergence towards a non-zero threshold distribution function (if \cnjref{rg-plop-threshold} is true) while the probability of \(\Pgiant(\beta)\) should converge to zero.
Similarly, for edge radius function \(\rn^2 = c/n\) we observe the opposite phenomenon: the probability of \(\Pgiant(\beta)\) begins to converge to a non-zero threshold distribution function (near \(c = 1.5\)) when \(\rn^2 = c/n\), while the probability of \(\Plopu\) converges to zero for all \(c\) at this choice of \(\rn^2\).

While we lack threshold distribution functions for both \(\Plopu\) and \(\Pgiant\), we include the established upper bound for \(\Plop\) (\prpref{rg-plop-upperbound}) for comparison and plot each numerical curve for increasing network sizes \(n = \{10^2,10^3,10^4\}\).
We note that the bound in \prpref{rg-plop-upperbound} forbids only vertex-induced complete graphs of order \(6\) (\(K_6\)) which is looser than \(\Plopu\) which forbids edge-induced cycles of lengths \(k \geq 6, k \neq 7\) from the set \(\Fmc\).
The combination of plots in \figref{rg-prop-vs-c} serve to demonstrate the mutual exclusion between \(\Plop\) and \(\Pgiant\) as \(n\rightarrow\infty\).

\begin{figure}[!t]
  \centering%
  \includegraphics[width=\columnwidth]{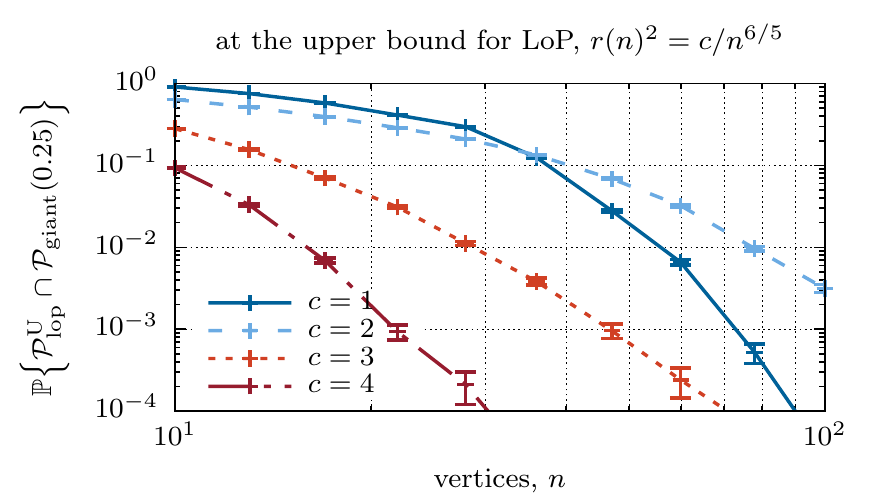}\\%
  \includegraphics[width=\columnwidth]{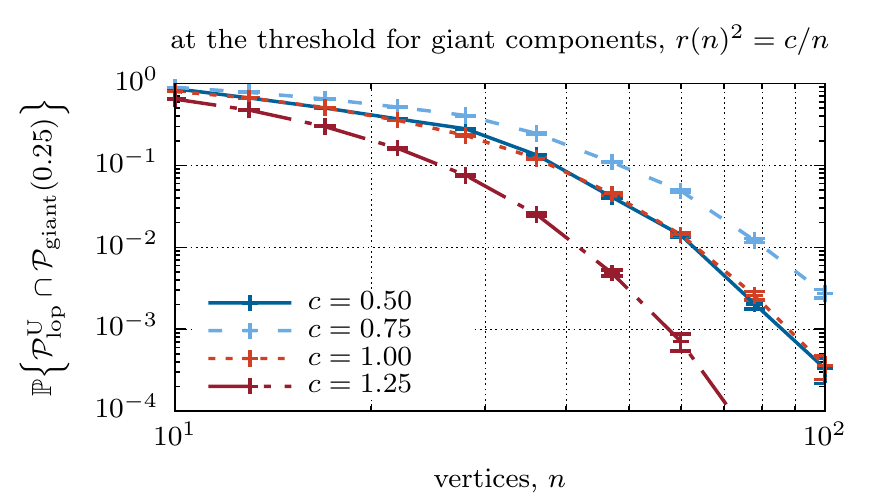}%
  \caption{%
    Numerical probability of satisfying both \(\Plopu\) and \(\Pgiant(0.25)\) in RG graphs plotted as a function of \(n\) where the edge radius is chosen according to \(\rn^2 = c/n^{6/5}\) (\textbf{top}) and \(\rn^2 = c/n\) (\textbf{bottom}).
    Numerical probabilities are computed with \(95\%\) confidence intervals generated from \(S = 10^5\) \iid graphs.
    \textsc{PlopU} was configured to use a maximum of \(I = 10^3\) iterations.
  }
  \label{fig:rg-prop-vs-n}
\end{figure}

In \figref{rg-prop-vs-n}, we focus on both edge radius functions selected for \figref{rg-prop-vs-c}, and instead parameterize by \(c\).
Appropriate parameter values are chosen to explore the area in the gaps presented in \figref{rg-prop-vs-c}.
For each edge radius function within this regime, we plot the probability that an RG graph satisfies both \(\Plopu\) and \(\Pgiant(0.25)\) as a function of the network size, \(n\).
We observe that the exclusion between \(\Plopu\) and \(\Pgiant\) develops even more quickly than in the case of ER graphs.
The increase in speed at which this exclusion develops is likely due to the separation in order between the thresholds functions that give rise to \(\Plop\) and \(\Pgiant\), which was not present in ER graphs.

\section{Conclusion}\label{sec:conclusions}

In this paper, we investigated the achievable fraction of the capacity region of Greedy Maximal Scheduling via an analytical tool known as Local Pooling.
We provided rigorous characterizations of the LoP factor in large networks modeled as \Erdos-\Renyi (ER) and random geometric (RG) graphs under the primary interference model.
We employed threshold functions to establish critical values for either the edge probability or communication radius to yield useful bounds on the range and expectation of the LoP factor as the network grows large in size.
For sufficiently dense random graphs, we found that the LoP factor is bounded between \(1/2\) and \(2/3\), while sufficiently sparse random graphs permit GMS optimality (the LoP factor is \(1\)) with high probability.
We observed that edge densities permitting connectivity generally admit cycle subgraphs which forms the basis for the LoP factor upper bound of \(2/3\) and concluded with simulations that explored this aspect.
In the regime of small network sizes, our simulation results suggest the probability that an ER or RG graph satisfies LoP and is connected decays rather quickly with the size of the network.

Avenues for future investigation of LoP include a more rigorous examination of the rate of convergence of the probabilities of these graph properties to their asymptotic values.
Additionally, examining the fraction of nodes/edges/components in the network satisfying LoP (\(\sigma = 1\)) may help in identifying simple topology control techniques to increase the LoP factor (\eg, removal of edges to break forbidden cycles in non-LoP satisfying components or addition of edges to patch smaller LoP-satisfying components together).

\appendix

\subsection{Ancillary Lemmas}

\lemref{er-exp-cycles} and \lemref{er-exp-dumbbells} are used to prove \thmref{er-plop-threshold} (\prfref{er-plop-threshold}), while \lemref{rg-exp-edges} and \lemref{rg-clt-edges} are used in the proof of \prpref{rg-pedge-threshold} (\prfref{rg-pedge-threshold}).

\begin{lemma}[Expected Forbidden Cycles in \(\Gnpn\)]\label{lem:er-exp-cycles}
  When \(\pn \sim c/n, c < 1\), the expected number of forbidden cycles of \(\Fmc\) in \(\Gnpn\) obeys:
  \begin{equation}
    \limninfty \; \sum_{\mathclap{\substack{6 \leq k \leq n, \\ k \neq 7}}} \Expect{G(C_k)} = -\log(\sqrt{1\!-\!c}) - \sum_{k \in \Kmc } \frac{c^k}{2k}.
  \end{equation}
\end{lemma}
\begin{IEEEproof}
  Given the choice of \(\pn\), it follows that:
  \begin{equation}\label{eq:pn-bounds}
    \forall \delta > 0, \exists n_{\delta} > 0: \frac{c-\delta}{n} \leq \pn \leq \frac{c+\delta}{n}, \forall n > n_{\delta}.
  \end{equation}

  The expected number of copies of a \(k\)-length cycle, \(C_k\) in \(\Gnpn\) can be expressed as a product between the number of possible unlabelled cycles and the probability that each forms the desired cycle, \(\Expect{G(C_k)} = n^{\underline{k}} / (2k) * p(n)^k\).
  Incorporating the bounds in \eqref{pn-bounds} yields:
  \begin{equation}
    \frac{n^{\underline{k}}}{2k} \frac{(c-\delta)^k}{n^k} \leq \Expect{G(C_k)} \leq \frac{n^{\underline{k}}}{2k} \frac{(c+\delta)^k}{n^k}.
  \end{equation}

  Next evaluate the following series when \(c < 1\) and \(\delta \in (0,1-c)\):
  \begin{align}
    \limninfty \sum_{k=1}^{n} \frac{n^{\underline{k}}}{n^k} \frac{(c+\delta)^k}{2k} &\overset{(a)}{=} \sum_{k=1}^{\infty} \frac{(c+\delta)^k}{2k} \label{eq:series} \\
    &\overset{(b)}{=} -\log(\sqrt{1-(c+\delta)}) \\
    &\overset{(c)}{\leq} -\log(\sqrt{1-c}) + \eps.
  \end{align}
  where we apply
  \((a)\) the monotone convergence theorem,
  \((b)\) series convergence when \(c+\delta < 1\), and
  \((c)\) continuity and monotonicity of \(\log(\sqrt{1-c})\) at \(c\).

  By a similar process on the lower bound series, and by controlling \(\eps\) by choice of \(\delta\), we establish:
  \begin{equation}
    \limninfty \; \sum_{\mathclap{\substack{6 \leq k \leq n, \\ k \neq 7}}} \Expect{G(C_k)} = -\log(\sqrt{1-c}) - \sum_{k \in \Kmc } \frac{c^k}{2k},
  \end{equation}
  where we subtract out a finite number of terms (\(\Kmc = \{1,2,3,4,5,7\}\)) that were originally included in \eqref{series} but do not correspond to forbidden cycle lengths.
\end{IEEEproof}

\begin{lemma}[Expected Forbidden Dumbbells in \(\Gnpn\)]\label{lem:er-exp-dumbbells}
  When \(\pn \sim c/n, c < 1\), the expected number of forbidden dumbbells of \(\Fmc\) in \(\Gnpn\) obeys:
  \begin{equation}
    \limninfty \sum_{k = 0}^{n} \left(\Expect{G(D_{k}^{5,5})} + \Expect{G(D_{k}^{5,7})} + \Expect{G(D_{k}^{7,7})}\right) = 0.
  \end{equation}
\end{lemma}
\begin{IEEEproof}
  Given \(\pn\), it follows that:
  \begin{equation}
    \forall \delta > 0, \exists n_{\delta} > 0: \pn \leq \frac{c+\delta}{n}, \forall n > n_{\delta}.
  \end{equation}

  The expected number of dumbbells, \(D_{k}^{s,t}\) (unions of cycles of lengths \(s\) and \(t\) joined by a \(k\)-edge path), assuming \(s \neq t\) and \(k \geq 1\) is:
  \begin{align}
    \Expect{G(D_{k}^{s,t})} &= \frac{n^{\underline{s}}}{2s} \frac{(n\!-\!s)^{\underline{t}}}{2t} s (n\!-\!s\!-\!t)^{\underline{k-1}} t \pn^{s+t+k} \nonumber \\
    &< \frac{(c+\delta)^{s+t+k}}{4n}. \label{eq:expected-k-dumbbells}
  \end{align}
  where there are \(n^{\underline{s}} / (2s)\) unlabelled cycles \(C_s\), \((n - s)^{\underline{t}} / (2t)\) unlabelled cycles \(C_t\) from the remaining \(n - s\) vertices, and \(s(n - s - t)^{\underline{k - 1}}t\) ways of connecting \(C_s\) to \(C_t\) with a \(k\)-edge path using the remaining \(n - s - t\) vertices.
  The probability that such a selection of vertices forms \(D_{k}^{s,t}\) is \(\pn^{p+q+k}\).

  In the event the path contains no edges, \(k\!=\!0\), then the cycles share a common vertex:
  \begin{equation}\label{eq:expected-0-dumbbells}
    \Expect{G(D_{0}^{s,t})} = \frac{n^{\underline{s}}}{2s} \frac{(n\!-\!s)^{\underline{t\!-\!1}}}{2} s \pn^{s+t} < \frac{(c+\delta)^{s+t}}{4n}
  \end{equation}
  In this case, \(C_t\) is created using one vertex from \(C_s\) and a \((t\!-\!1)\)-edge path from the remaining \(n\!-\!s\) vertices.

  Finally, if \(s\!=\!t\), then \(\Expect{G(D_{k}^{s,t})}\) contains an additional factor of \(1/2\) due to symmetry, but nevertheless is upper bounded by the expressions in \eqref{expected-k-dumbbells} and \eqref{expected-0-dumbbells}.

  It remains to show that expected number of all forbidden dumbbells is zero.
  Let \(\hat{c} = c+\delta < 1\) for an appropriate choice of \(\delta \in (0,1-c)\):
  \begin{align}
    \limninfty &\sum_{k = 0}^{n} \left(\Expect{G(D_{k}^{5,5})} + \Expect{G(D_{k}^{5,7})} + \Expect{G(D_{k}^{7,7})}\right) \\
    &\overset{(a)}{<} \limninfty \frac{\hat{c}^{10} + \hat{c}^{12} + \hat{c}^{14}}{4n} \sum_{k=0}^{\infty} \hat{c}^k \overset{(b)}{=} 0,
  \end{align}
  where we
  \((a)\) apply bounds derived above and collect common factors, and
  \((b)\) apply geometric series convergence and evaluate the limit.
\end{IEEEproof}

\begin{lemma}[Expected Edges in \(\Gnrn\)]\label{lem:rg-exp-edges}
  If \(\rn^2 \in 2c / (\pi n) + x2\sqrt{cn} / (\pi n^2)\) with \(x \in \Rbb\), then the mean number of edges \(\Mnrn\) in \(\Gnrn\) is:
  \begin{equation}
    \Expect{\Mnrn} \in cn + x\sqrt{cn} + \oh{\sqrt{n}}.
  \end{equation}
\end{lemma}
\begin{IEEEproof}
  This follows from a specialization of \cite[Prop.~3.1]{Pen2003} which provides the asymptotic mean of a subgraph count of \(\Gnrn\) when \(\rn \in \oh{1}\).
  We will not recreate the theory here, but instead provide enough direction to allow the reader to follow along with \cite{Pen2003}.
  The expected number of edges (\(\Expect{\Mnrn}\)) is given as:
  \begin{equation}
    \Expect{\Mnrn} \sim \mu_{K_2,\Rbb^2}\rn^{d(k-1)}n^k,
  \end{equation}
  where the subgraph \(K_2\) (the complete graph on \(2\) vertices, \ie, an edge) has \(k=2\) vertices, \(d=2\) is the dimension of the space in which the points of \(\Gnrn\) reside, and \(\mu_{K_2,\Rbb^2}\) is computed as follows:
  \begin{align}
    \mu_{K_2,\Rbb^2} &\stackrel{(a)}{=} \frac{1}{2!} \int_{\mathrlap{\Rbb^2}} f(x)^2 \drm x \int_{\mathrlap{\Rbb^2}} h_{K_2}(\{0,x_1\}) \drm x_1 \\
    &\stackrel{(b)}{=} \frac{1}{2} \int_{\mathrlap{\Rbb^2}} h_{K_2}(\{0,x_1\})\drm x_1 \stackrel{(c)}{=} \frac{\pi}{2}.
  \end{align}
  where
  \((a)\) is simplified from \cite{Pen2003} for the subgraph type \(K_2\),
  \((b)\) follows from \(f(x)\) being the uniform distribution over the unit square \([-1/2,1/2]^2\) used to generate \iid vertex positions, and
  \((c)\) follows from \(h_{K_2}(0,x_1)\) being the indicator function on whether or not two vertices (one at the origin and the other at \(x_1\)) with unit edge distance form \(K_2\).
  To form \(K_2\), \(x_1\) must be within the unit disk centered at the origin to connect to the vertex at the origin.

  Finally, expanding \(\mu_{K_2}\rn^2n^2\) and grouping \(\oh{\sqrt{n}}\) terms is sufficient.
\end{IEEEproof}

\begin{lemma}[CLT for Edges in \(\Gnrn\)]\label{lem:rg-clt-edges}
  If \(\rn^2 \in 2c / (\pi n) + 2\sqrt{cn} x / (\pi n^2)\) with \(x \in \Rbb\), then the centered and scaled number of edges \(\Mnrn\) in \(\Gnrn\) converges in distribution to that of a centered normal \rv{}:
  \begin{equation}
    \frac{\Mnrn - \Expect{\Mnrn}}{n^{1/2}} \overset{\Dmc}{\sim} \Nmc(0,c).
  \end{equation}
\end{lemma}
\begin{IEEEproof}
  This follows from a specialization of \cite[Thm.~3.13]{Pen2003} which provides a central limit theorem for collections of subgraph counts of \(\Gnrn\) when \(\limninfty n\rn^d \rightarrow \rho \in (0,\infty)\).
  The distribution of the centered and scaled number of edges \((\Mnrn - \Expect{\Mnrn}) / \sqrt{n}\) is an asymptotic centered normal with variance:
  \begin{equation}\label{eq:rg-edge-var}
    \left(\sum_{j=1}^{k} \rho^{2k-j-1} \Phi_j(K_2,K_2) \right) - k^2\rho^{2k-2}\mu_{K_2}^2,
  \end{equation}
  where the subgraph \(K_2\) (the complete graph on \(2\) vertices, \ie, an edge) has \(k=2\) vertices, \(d=2\) is the dimension of the space in which the points of \(\Gnrn\) reside, \(\mu_{K_2,\Rbb^2} = \pi/2\) is computed as shown in the proof of \lemref{rg-exp-edges}, and \(\Phi_1(K_2,K_2)\) simplifies to:
  \begin{equation}
    \Phi_1(K_2,K_2) = \!\int_{\mathrlap{\Rbb^2}} h_{K_2}(0,x_2) \drm x_2 \!\int_{\mathrlap{\Rbb^2}} h_{K_2}(0,x_3) \drm x_3 = \pi^2,
  \end{equation}
  and \(\Phi_2(K_2,K_2) = \mu_{K_2}\).

  Finally, note that for the given \(\rn^2\), \(n\rn^2 \sim \rho = 2c / \pi\).
  Substituting \(\Phi_1\), \(\Phi_2\), \(\mu_{K_2}\), and \(\rho\) into \eqref{rg-edge-var}, we obtain the asymptotic variance:
  \begin{equation}
    \rho^2\Phi_1(K_2,K_2) + \rho\Phi_2(K_2,K_2) - 4\rho^2\mu_{K_2}^2 = \rho\mu_{K_2} = c.
  \end{equation}
\end{IEEEproof}

\subsection{%
  \lemref{plop-monotonicity} (\texorpdfstring{\(\Plop\)}{Plop} Monotonicity)%
}\label{prf:plop-monotonicity}

\begin{IEEEproof}
  Let \(G \in \Plop\).
  From \thmref{plop}, \(G\) contains no edge-induced forbidden subgraphs from \(\Fmc\).
  Let \(H \subset G\) by an appropriate removal of edges.
  The removal of edges from \(G\) cannot possibly create edge-induced forbidden subgraphs where none existed before, therefore \(H \in \Plop\) and \(\Plop\) is monotone decreasing as described by \defref{monotonic-property}.
\end{IEEEproof}

\subsection{%
  \lemref{plop-prop-bounds} (Separate Sufficient and Necessary Cond.\ for \texorpdfstring{\(\Plop\)}{Plop})%
}\label{prf:plop-prop-bounds}

\begin{IEEEproof}
  Since all forbidden subgraphs in \(\Fmc\) (\thmref{plop}) contain cycles, it immediately follows that forbidding all cycles (\(\Plopl\)) is sufficient for \(\Plop\).
  Separately, forbidding any subset of subgraphs in \(\Fmc\) is a necessary condition for \(\Plop\), therefore forbidding cycles of lengths \(k \geq 6, k \neq 7\) (\(\Plopu\)) is necessary for \(\Plop\).
  Thus, the subsets of graphs on \(n\) vertices that satisfy \(\Plopl\), \(\Plop\), \(\Plopu\) can be nested in that order.
\end{IEEEproof}

\subsection{%
  \lemref{plop-prob-bounds} (Probability Bounds for \texorpdfstring{\(\Plop\)}{Plop})%
}\label{prf:plop-prob-bounds}

\begin{IEEEproof}
  Given \(\pn\) (or \(\rn\)) and \(n \in \Zbb^+\), \(\Gnpn\) (or \(\Gnrn\)) is a random graph generated from a distribution on \(\GGn\).
  Interpreted as events, the nesting of subsets \(\Plopl\), \(\Plop\), \(\Plopu\) by \lemref{plop-prop-bounds} provides the desired ordering of probabilities.
\end{IEEEproof}

\subsection{%
  \prpref{er-pedge-threshold} (Reg.\ Sharp Threshold for \texorpdfstring{\(\Pedge\)}{Pedge} in \texorpdfstring{\(\Gnpn\)}{Gnpn})%
}\label{prf:er-pedge-threshold}

\begin{IEEEproof}
  Let the \rv{} \(\Mnpn\) (shortened to \(M\)) be the number of edges in graph \(\Gnpn\).
  We show for the given choice of \((\psn = 2c / n, \an = 2\sqrt{cn} / n^2)\) and \(F(x) = \Phi(-x)\) that:
  \begin{equation}
    \pn \sim \psn + x\an \Rightarrow \limninfty \Prob{M \leq cn} = F(x),
  \end{equation}
  holds for every point of continuity of \(F(x)\), \(x \in \Rbb\).

  \(\Mnpn\) has a binomial \pdf{}; for \(\pn \sim \psn + x\an\), \(\Mnpn\) has mean and variance:
  \begin{align}
    \Expect{\Mnpn} &= \binom{n}{2}\pn = cn + x\sqrt{cn} + \oh{\sqrt{n}} \label{eq:er-edge-mean} \\
    \Var{\Mnpn} &= \binom{n}{2}\pn(1-\pn) = cn + \oh{n},                \label{eq:er-edge-variance}
  \end{align}
  by using the additional facts \(\pn = 2c / n + \oh{1/n}\) and \(\pn^2 = \oh{1/n}\).

  Finally, for \(\pn \sim \psn + x\an\):
  \begin{align}
    \Prob{M \leq cn} &\stackrel{(a)}{=} \Prob{\frac{M - \Expect{M}}{\sqrt{\Var{M}}} \leq \frac{cn - \Expect{M}}{\sqrt{\Var{M}}}} \\
    &\stackrel{(b)}{=} \Prob{\frac{M - \Expect{M}}{\sqrt{\Var{M}}} \leq \frac{-x\sqrt{cn} + \oh{\sqrt{n}}}{\sqrt{cn + \oh{n}}}} \\
    &\stackrel{(c)}{=} \Phi\left(-x + \oh{1}\right) + \oh{1} \\
    &\stackrel{(d)}{=} \Phi(-x) + \oh{1},
  \end{align}
  where we
  \((a)\) standardize \(\Mnpn\),
  \((b)\) expand using \eqref{er-edge-mean} and \eqref{er-edge-variance},
  \((c)\) asymptotically simplify the inequality's \rhs and apply the CLT to the standardized \(\Mnpn\), and
  \((d)\) apply continuity of the standard normal \cdf, \(\Phi(x)\).

  Thus, for the specific case when \(c=2\), we conclude:
  \begin{equation}
    \limninfty \Prob{\Gnpn \in \Pedge} = \limninfty \Prob{M \leq 2n} = \Phi\left(-x\right).
  \end{equation}
\end{IEEEproof}

\subsection{%
  \thmref{er-plop-threshold} (Reg.\ Threshold for \texorpdfstring{\(\Plop\)}{Plop} in \texorpdfstring{\(\Gnpn\)}{Gnpn})%
}\label{prf:er-plop-threshold}

\begin{IEEEproof}
  Let \(\Gamma\) be a connected graph.
  Let the \rv \(G(\Gamma)\) be the number of copies of \(\Gamma\) in graph \(\Gnpn\).
  Let \(A_{\Gamma} = \{G(\Gamma) > 0\}\) be the event that there are one or more copies of \(\Gamma\) in \(\Gnpn\).
  We show for the given \(\psn\) and \(F(x)\), that:
  \begin{equation}
    \pn \sim x\psn \Rightarrow \limninfty \Prob{\Plop} = F(x),
  \end{equation}
  holds for every point of continuity of \(F(x)\), \(x \in \Rbb\).

  Suppose \(\pn \sim x\psn\).
  We first upper bound \(\Prob{\Plop}\):
  \begin{equation}
    \limninfty \Prob{\Plop} \overset{(a)}{\leq} \limninfty \Prob{ \bigcap_{\mathrlap{\!\!\!\substack{6 \leq k \leq K, \\k \neq 7}}} \overline{A_{C_k}} }
    \overset{(b)}{=} \exp\!\left(\!-\sum_{\mathclap{\!\substack{6 \leq k \leq K, \\ k \neq 7}}} \; \frac{x^k}{2k}\!\right), \label{eq:plop-upper-bound}
  \end{equation}
  where
  \((a)\) follows by forbidding only cycles in \(\Fmc\) up to length \(K \leq n\), and
  \((b)\) is a consequence of \cite[Cor.~4.9]{Bol2001} which shows that when \(\pn \sim x/n\), a finite-length random vector of cycle subgraph counts \(\{G(C_k)\}\) converges in distribution to that of independent Poisson \rv's with means \(\{\lambda_k = x^k / (2k)\}\).

  Now, considering the upper bound, suppose \(x \geq 1\).
  The series \(\sum_{k=1}^{\infty} x^k / (2k)\) diverges to \(\infty\), thus \(\Prob{\Plop}\) can be upper-bounded by arbitrarily small \(\eps\) by a large enough choice of \(K\).
  Note when \(x = 1\), the series becomes the harmonic series, which also diverges, albeit more slowly.
  Thus,
  \begin{equation}
    \pn \sim x\psn, x \geq 1 \Rightarrow \limninfty \Prob{\Plop} = 0.
  \end{equation}

  Alternatively, consider the upper bound when \(x < 1\).
  The series \(\sum_{k=1}^{\infty} x^k / (2k)\) converges to \(-\log(\sqrt{1-x})\).
  Thus, for arbitrarily small \(\eps\), a sufficiently large choice for \(K\) will yield:
  \begin{equation}\label{eq:series-bound}
    -\sum_{\substack{6 \leq k \leq K, \\ k \neq 7}} \frac{x^k}{2k} \leq \log(\sqrt{1-x}) + \sum_{k \in \Kmc} \frac{x^k}{2k} + \eps
  \end{equation}
  where \(\Kmc = \{1,2,3,4,5,7\}\).
  Substituting \eqref{series-bound} into \eqref{plop-upper-bound}, we obtain the following upper bound for \(\limninfty \Prob{\Plop}\):
  \begin{equation}\label{eq:plop-upper-bound2}
    \limninfty \Prob{\Plop} < \sqrt{1-x} \exp\left(\sum_{k \in \Kmc} \frac{x^k}{2k}\right) + \eps',
  \end{equation}
  where \(\exp(\eps) \leq 1 + (e-1)\eps\) and the constants in front of \(\eps\) can be rolled into \(\eps' > 0\).

  It remains to provide a lower bound when \(x < 1\).
  We start by lower bounding \(\Prob{\Plop}\):
  \begin{align}
    \Prob{\Plop} &= \Prob{ \bigcap_{\Gamma \in \Fmc} \overline{A_{\Gamma}} } \overset{(a)}{\geq} \prod_{\Gamma \in \Fmc} \Prob{\overline{A_{\Gamma}}} \\
    &\overset{(b)}{\geq} \prod_{\Gamma \in \Fmc} \exp\left(-\frac{\Expect{G(\Gamma)}}{1-\pn}\right) \\
    &= \exp\left(-\frac{1}{1-\pn} \sum_{\Gamma \in \Fmc} \Expect{G(\Gamma)} \right) \label{eq:plop-lower-bound}
  \end{align}
  where
  \((a)\) follows from the FKG Inequality applied to the set of monotone decreasing properties \(\overline{A_{\Gamma}}\) on \(\GGnpn\) \cite[Thm.~2.12]{JanLucRuc2000}, and
  \((b)\) is the result of applying \cite[Cor.~2.13]{JanLucRuc2000} to each multiplicand to obtain an exponential lower bound.

  First, we note that:
  \begin{equation}\label{eq:limit-pnfrac}
    \limninfty \frac{1}{1-\pn} = 1
  \end{equation}

  Second, by \lemref{er-exp-cycles} and \lemref{er-exp-dumbbells} (with \(\pn \sim x/n, x < 1\)), the limit of the sum of the expected forbidden subgraph counts depends solely on cycles:
  \begin{equation}\label{eq:limit-subgraphs}
    \limninfty \sum_{\Gamma \in \Fmc} \Expect{G(\Gamma)} = -\log(\sqrt{1-x}) - \sum_{k \in \Kmc} \frac{x^k}{2k},
  \end{equation}
  with \(\Kmc = \{1,2,3,4,5,7\}\).

  Thus, by making use of \eqref{limit-pnfrac} and \eqref{limit-subgraphs} in \eqref{plop-lower-bound}, the limiting probability of satisfying \(\Plop\) is lower bounded by:
  \begin{equation}\label{eq:plop-lower-bound2}
    \limninfty \Prob{\Plop} \geq \sqrt{1-x} \exp\left(\sum_{k \in \Kmc} \frac{x^k}{2k}\right)
  \end{equation}

  Finally, combining \eqref{plop-upper-bound2} and \eqref{plop-lower-bound2} produces our desired limit when \(x<1\).
\end{IEEEproof}

\subsection{%
  \corref{er-plop-threshold-weak} (Threshold Function for \texorpdfstring{\(\Plop\)}{Plop} in \texorpdfstring{\(\Gnpn\)}{Gnpn})%
}\label{prf:er-plop-threshold-weak}

\begin{IEEEproof}
  This, follows directly from the \thmref{er-plop-threshold} and the monotonicity of property \(\Plop\).
  If \(\pn \in \om{\psn}\), there exists \(x > 1\) for which \(\pn\) is asymptotically greater than \(x/n\).
  Alternately, if \(\pn \in \oh{\psn}\), then \(\pn\) is asymptotically less than \(x/n\) for all \(x > 0\).
\end{IEEEproof}

\subsection{%
  \prpref{er-sigma-prob-bounds} (\texorpdfstring{\(\sigma\)-LoP}{LoP Factor} Bounds in \texorpdfstring{\(\Gnpn\)}{Gnpn})
}\label{prf:er-sigma-prob-bounds}

\begin{IEEEproof}
  Let \(G = \Gnpn\).
  Let \(\Gamma\) and \(A_{\Gamma}\) be as defined in \prfref{er-plop-threshold}.
  \begin{align}
    \limninfty &\Prob{1/2 \leq \sigma(G) \leq 2/3} \stackrel{(a)}{=} \limninfty \Prob{\sigma(G) \leq 2/3} \\
    &\stackrel{(b)}{\geq} 1 - \limninfty \Prob{\bigcap_{k=1}^{\infty} \overline{A_{C_{6k}}}} \stackrel{(c)}{=} 1,
  \end{align}
  where
  \((a)\) the lower bound is always true by \lemref{sigma-bounds},
  \((b)\) the presence of any \(C_{6k}\) is sufficient for \(\sigma(G) \leq 2/3\) by \lemref{sigma-C6k}, and
  \((c)\) \(\pn\) is above the joint threshold for the appearance of all \(C_{6k}\) by appropriate `thinning' of the argument of \thmref{er-plop-threshold} in \prfref{er-plop-threshold} to \(C_{6k}\) and divergence of the series for \(c > 1\).
\end{IEEEproof}

\subsection{%
  \thmref{er-sigma-exp-bounds} (\texorpdfstring{\(\Expect{\sigma}\)}{Expected LoP Factor} Bounds in \texorpdfstring{\(\Gnpn\)}{Gnpn})
}\label{prf:er-sigma-exp-bounds}

\begin{IEEEproof}
  For convenience, let \(G = \Gnpn\).
  We first consider the lower bound:
  \begin{align}
    \Expect{\sigma(G)}
    &= \Expect{\sigma(G) | G \in \Plop}\Prob{G \in \Plop} \nonumber \\
    &\qquad + \Expect{\sigma(G) | G \notin \Plop} \Prob{G \notin \Plop} \\
    &\stackrel{(a)}{=} \Prob{G \in \Plop} \nonumber \\
    &\qquad + \Expect{\sigma(G) | G \notin \Plop} (1 - \Prob{G \in \Plop}) \\
    &\stackrel{(b)}{\geq} \Prob{G \in \Plop} + \frac{1}{2} (1 - \Prob{G \in \Plop}) \\
    &= \frac{1}{2}\left(1 + \Prob{G \in \Plop}\right),
  \end{align}
  where
  \((a)\) \(\sigma(G)=1\) when \(G\in\Plop\), and
  \((b)\) \(\sigma(G)\geq 1/2\) when \(G\notin\Plop\).
  Finally, take the limit as \(n\rightarrow\infty\) and apply \(\Prob{G\in\Plop} \rightarrow F_l(x)\) from \thmref{er-plop-threshold}.

  We now apply a similar argument to the upper bound, but partition on the presence of the class of cycles \(\{C_{6k},k\in\Nbb_+\}\), all of which result in \(\sigma \leq 2/3\).
  Let \(\Gamma\) and \(A_{\Gamma}\) be as defined in \prfref{er-plop-threshold}, and let \(\overline{\Amc} \equiv \{\cap_{k=1}^{\infty} \overline{A_{C_{6k}}}\}\) be the event that there exist no cycles \(C_{6k}\) within \(G\):
  \begin{align}
    &\Expect{\sigma(G)} = \Expect{\sigma(G) | \Amc} \Prob{\Amc} + \Expect{\sigma(G) | \overline{\Amc}} \Prob{\overline{\Amc}} \\
    &\qquad \stackrel{(a)}{\leq} \frac{2}{3} (1 - \Prob{\overline{\Amc}}) + \Prob{\overline{\Amc}} = \frac{1}{3}\left(2 + \Prob{\overline{\Amc}}\right),
  \end{align}
  where \((a)\) \(\sigma(G) \leq 2/3\) when \(\Amc\), and \(\sigma(G) \leq 1\) is always true.
  Evaluating the limit of \(\Prob{\overline{\Amc}}\) can be done using the same approach as the argument of \thmref{er-plop-threshold} in \prfref{er-plop-threshold}:
  \begin{align}
    \limninfty &\Prob{ \bigcap_{k=1}^{K} C_{6k} \nsubseteq G} = \prod_{k=1}^{K} \exp\left(-\frac{x^{6k}}{12k} \right) \\
    &= \exp\left( -\sum_{k=1}^{K} \frac{x^{6k}}{12k} \right) \\
    &= \exp\left(\frac{1}{12}\log(1-x^6)\right) + \eps_K \\
    &= (1-x^6)^{1/12} + \eps_K,
  \end{align}
  when \(x \leq 1\) where \(\eps_K\) can be driven lower by a larger choice of \(K\).
  Otherwise, when \(x > 1\), the series in the exponent diverges (\ie, \(G\) is sure to contain a cycle in \(\{C_{6k}\}\)).
\end{IEEEproof}

\subsection{%
  \corref{er-pgiant-threshold} (Reg.\ Threshold for \texorpdfstring{\(\Pgiant(\beta)\)}{Pgiant(B)} in \texorpdfstring{\(\Gnpn\)}{Gnpn})%
}\label{prf:er-pgiant-threshold}

\begin{IEEEproof}
  Given \(\beta^* \in (0,1)\), construct \(\psn = c(\beta^*)/n\) using \eqref{c-beta}.
  We show that:
  \begin{equation}
    \pn \sim x \psn \Rightarrow \limninfty \Prob{\Gnpn \in \Pgiant(\beta^*)} = F(x),
  \end{equation}
  where \(F(x) = \mathbf{1}\{x > 1\}\) for all continuity points of \(F(x)\): \(\Rbb\setminus\{1\}\).

  Suppose \(\pn \sim x\psn\), with \(x > 1\).
  \(\pn\) is asymptotically larger than \(\psn\) and by monotonicity of \eqref{c-beta}, there exists \(\beta \in (\beta^*,1)\) such that:
  \begin{equation}
    \exists n_0 > 0, \forall n > n_0 : p(n) > \frac{c(\beta)}{n} > \frac{c(\beta^*)}{n}.
  \end{equation}
  Apply part \emph{ii)} of \cite[Thm.~5.4]{JanLucRuc2000} to establish that the size of the largest component, denoted as \(\Lnpn\), converges in probability to \(\beta n\):
  \begin{equation}
    \forall \eps > 0, \limninfty \Prob{\left|\frac{\Lnpn}{\beta n} - 1\right| < \eps} = 1.
  \end{equation}
  By choosing \(\eps\) such that \(\beta^* = (1-\eps)\beta\), the event \(\left| \Lnpn / (\beta n) - 1\right| < \eps\) is a subset of the event that \(\Lnpn \geq \beta^*n\), giving us the upper bound:
  \begin{equation}
    \Prob{\left|\frac{\Lnpn}{\beta n} - 1\right| < \eps} \leq \Prob{\frac{\Lnpn}{n} \geq \beta^*}.
  \end{equation}
  Since \(\limninfty \Prob{\left| \Lnpn / (\beta n) - 1\right| < \eps} = 1\) and probabilities are bounded above by \(1\), we apply the squeeze theorem and conclude that \(\limninfty \Prob{ \Lnpn / n > \beta^*} = 1\).

  Alternately, suppose \(\pn \sim x\psn\), with \(x < 1\).
  \(\pn\) is asymptotically smaller than \(\psn\) and by monotonicity of \eqref{c-beta}, there exists \(\beta \in (0,\beta^*)\) such that:
  \begin{equation}
    \exists n_0 > 0, \forall n > n_0 : p(n) < \frac{c(\beta)}{n} < \frac{c(\beta^*)}{n}.
  \end{equation}
  Again, we use \cite[Thm.~5.4]{JanLucRuc2000} to show that \(\Lnpn\) converges in probability to \(\beta n\):
  \begin{equation}
    \forall \eps > 0, \limninfty \Prob{\left|\frac{\Lnpn}{\beta n} - 1\right| < \eps} = 1.
  \end{equation}
  By choosing \(\eps\) such that \((1+\eps)\beta = \beta^*\), the event \(\left| \Lnpn / (\beta n) - 1\right| < \eps\) is a subset of the event that \(\Lnpn < \beta^*n\), giving us the upper bound:
  \begin{equation}
    \Prob{\left|\frac{\Lnpn}{\beta n} - 1\right| < \eps} \leq 1 - \Prob{\frac{\Lnpn}{n} \geq \beta^*}
  \end{equation}

  Since \(\limninfty \Prob{\left| \Lnpn / (\beta n) - \right| < \eps} = 1\) and probabilities are bounded below by \(0\), we apply the squeeze theorem and conclude that \(\limninfty \Prob{ \Lnpn / n \geq \beta^*} = 0\).
\end{IEEEproof}

\subsection{%
  \thmref{er-exclusion-plop-pgiant} (Mutual Excl.\ of \texorpdfstring{\(\Plop\)}{Plop} and \texorpdfstring{\(\Pgiant(\beta)\)}{Pgiant(B)} in \texorpdfstring{\(\Gnpn\)}{Gnpn})%
}\label{prf:er-exclusion-plop-pgiant}

\begin{IEEEproof}
  By \thmref{er-plop-threshold}, \(\pn \sim c/n, c>1\) implies that \(\Plop\) holds \aan.
  Therefore, \(\pn \sim c/n, c \leq 1\) is a necessary condition for \(\Plop\) to hold \aas.
  Under this necessary condition, we see that \(\pn\) is asymptotically less than \(c(\beta)/n\) since \(c(\beta) > 1\) and by \corref{er-pgiant-threshold}, \(\Pgiant(\beta)\) holds \aan.

  Thus, for \(\pn \sim c/n, \forall c \leq 1\):
  \begin{equation}
    0 \leq \limninfty \Prob{\Pgiant(\beta) \cap \Plop} \leq \limninfty \Prob{\Pgiant(\beta)} = 0
  \end{equation}

  Alternately, for \(\pn \sim c/n, \forall c > 1\):
  \begin{equation}
    0 \leq \limninfty \Prob{\Pgiant(\beta) \cap \Plop} \leq \limninfty \Prob{\Plop} = 0
  \end{equation}

  In both cases, we can conclude that \(\limninfty \Prob{\Pgiant(\beta) \cap \Plop} = 0\).
\end{IEEEproof}

\subsection{%
  \prpref{rg-pedge-threshold} (Reg.\ Sharp Threshold for \texorpdfstring{\(\Pedge\)}{Pedge} in \texorpdfstring{\(\Gnrn\)}{Gnrn})%
}\label{prf:rg-pedge-threshold}

\begin{IEEEproof}
  Let the \rv{} \(\Mnrn\) (shortened to \(M\)) be the number of edges in graph \(\Gnrn\).
  We show that for the given choice of \((\rsn^2 = 2c / (\pi n), \an = 2\sqrt{cn} / (\pi n^2)\) and \(F(x) = \Phi(-x)\) that:
  \begin{equation}
    \rn^2 \sim \rsn^2 + x\an \Rightarrow \limninfty \Prob{M \leq cn} = F(x),
  \end{equation}
  where \(F(x) = \Phi(-x)\) for all continuous points of \(F(x)\): \(\Rbb\).

  For \(\rn^2 \sim \rsn^2 + x\an\):
  \begin{align}
    \Prob{M \leq cn} &\overset{(a)}{=} \Prob{\frac{M - \Expect{M}}{\sqrt{n}} \leq \frac{cn - \Expect{M}}{\sqrt{n}}} \\
    &\overset{(b)}{=} \Prob{\frac{M - \Expect{M}}{\sqrt{n}} \leq -x\sqrt{c} + \oh{1}} \\
    &\overset{(c)}{=} \Phi\left(\frac{-x\sqrt{c} + \oh{1}}{\sqrt{c}}\right) + \oh{1} \\
    &\overset{(d)}{=} \Phi\left(-x + \oh{1}\right) + \oh{1} \\
    &\overset{(e)}{=} \Phi\left(-x\right) + \oh{1},
  \end{align}
  where we
  \((a)\) standardize \(\Mnrn\),
  \((b)\) apply \lemref{rg-exp-edges},
  \((c)\) apply \lemref{rg-clt-edges} and standardize the argument to the \cdf,
  \((d)\) results from asymptotic simplification, and
  \((e)\) apply continuity of the standard normal \cdf, \(\Phi(x)\).

  Thus, for the specific case when \(c=2\), we conclude:
  \begin{equation}
    \limninfty \Prob{\Gnrn \in \Pedge} = \limninfty \Prob{M \leq 2n} = \Phi\left(-x\right).
  \end{equation}
\end{IEEEproof}

\subsection{%
  \prpref{rg-plop-upperbound} (Upper Bound for \texorpdfstring{\(\Plop\)}{Plop} in \texorpdfstring{\(\Gnrn\)}{Gnrn})%
}\label{prf:rg-plop-upperbound}

\begin{IEEEproof}
  Let \(\Gamma_k\) be a feasible, connected, order \(k\) graph.
  Let \(G_e(\Gamma_k)\) and \(G_v(\Gamma_k)\) be the edge-induced and vertex-induced subgraph counts of \(\Gamma_k\) on graph \(\Gnrn\), resp.
  Let \(A_{\Gamma_k} = \{G_e(\Gamma_k) \geq 1\}\) and \(B_{\Gamma_k} = \{G_v(\Gamma_k) \geq 1\}\) be the events that there are one or more edge-induced or vertex-induced copies of \(\Gamma_k\) in \(\Gnrn\), resp.

  A necessary condition for \(\Plop\) is the absence of edge-induced cycles of length \(6\), which can be expressed as an intersection of a finite number of vertex-induced events, \(\{\overline{B_{\Gamma_6}}\}\):
  \begin{equation}
    \Prob{\Plop} \leq \Prob{\overline{A_{C_6}}} = \Prob{\cap_{\Gamma_6 \in \Ymc} \overline{B_{\Gamma_6}}},
  \end{equation}
  where \(\Ymc \equiv \{\Gamma_6 : \Gamma_6 \subseteq K_6, C_6 \subseteq \Gamma_6, \Gamma_6 \textup{ feasible}\}\).

  By \cite[Thm.~3.5]{Pen2003}, the finite collection of vertex-induced subgraph counts \(\{G_v(\Gamma_6)\}\) converge to independent Poisson \rv's with rates \(\{\lambda = c^5\mu_{\Gamma_6}\}\), for our choice of \(\rn^2\).
  The null probability of the subgraph counts becomes:
  \begin{align}
    \!\!\!\limninfty \Prob{\bigcap_{\mathrlap{\!\!\!\Gamma_6 \in \Ymc}} \overline{B_{\Gamma_6}}}
    = \prod_{\mathclap{\Gamma_6 \in \Ymc}} e^{-c^5 \mu_{\Gamma_6}}
    = \exp\!\left(\!\!-c^5 \sum_{\mathclap{\Gamma_6 \in \Ymc}} \mu_{\Gamma_6}\!\!\right),
  \end{align}
  where \(\mu_{\Gamma_6}\) is computed from \cite[Eq.~3.2]{Pen2003} for each vertex-induced subgraph.
  We may upper bound the exponential by considering a single term in the summation where \(\Gamma_6 = K_6\) (the complete graph on \(6\) vertices) has \(k=6\) vertices and then expressing a lower bound for \(\mu_{K_6,\Rbb^2}\):
  \begin{align}
    &\!\!\! \mu_{K_6,\Rbb^2} \!\stackrel{(a)}{=}\! \frac{1}{6!} \!\! \int_{\mathrlap{\Rbb^2}} f(x)^2 \drm x \!\! \int_{\mathrlap{\Rbb^2}} h_{K_6}(\{0,x_1,...,x_5\}) \drm x_1,...,\drm x_5 \\
    &\quad \stackrel{(b)}{=}\! \frac{1}{6!} \! \int_{\mathrlap{\Rbb^2}} h_{K_6}(\{0,x_1,...,x_5\})\drm x_1,...,\drm x_5
    \stackrel{(c)}{\geq} \frac{(\pi/4)^5}{6!},
  \end{align}
  where \((a)\) is simplified from \cite[Eq.~3.2]{Pen2003} for the subgraph type \(K_6\), \((b)\) follows from \(f(x)\) being the uniform distribution over the unit square \([-1/2,1/2]^2\) used to generate \iid{} vertex positions, and \((c)\) follows from \(h_{K_6}(0,x_1,\dots,x_5)\) being the indicator function on whether or not six vertices (one fixed at the origin) with unit edge distance form \(K_6\).
  The lower bound results when limiting the placement of all five vertices to a disk of radius \(1/2\) centered at the origin.
  Under this assumption, all six vertices are connected, form \(K_6\), and yield \(h_{K_6}(0,x_1,\dots,x_5) = 1\).
\end{IEEEproof}

\subsection{%
  \lemref{rg-pconn-threshold} (Reg.\ Sharp Threshold for \texorpdfstring{\(\Pconn\)}{Pconn} in \texorpdfstring{\(\Gnrn\)}{Gnrn})%
}\label{prf:rg-pconn-threshold}

\begin{IEEEproof}
  Let \rv \(T = \Tnrn\) be the minimum edge distance that yields a connected graph for \(\Gnrn\).
  Thus, the graph \(\Gnrn\) is connected iff \(T \leq \rn\).
  Using a specialization of \cite[Cor.~13.21]{Pen2003}, we show that:
  \begin{equation}
    \rn^2 \sim \rsn^2 + x\an \Rightarrow \limninfty \Prob{T \leq \rn} = F(x),
  \end{equation}
  where \(F(x) = \erm^{-\erm^{-x}}\) for all continuous points of \(F(x)\): \(\Rbb\).

  For \(\rn^2 \in \rsn^2 + x\an + \oh{\an}\):
  \begin{align}
    \Prob{T \leq \rn} &\stackrel{(a)}{=} \limninfty \Prob{n\pi T^2 - \log(n) \leq x + \oh{1}} \\
    &\stackrel{(b)}{=} \erm^{-\erm^{-x + \oh{1}}} + \oh{1} \stackrel{(c)}{=} \erm^{-\erm^{-x}} + \oh{1}
  \end{align}
  where
  \((a)\) follows from squaring both sides of the inequality and expanding \(\rn^2\) in terms of the given \(\rsn^2\) and \(\an\),
  \((b)\) results from a specialization of \cite[Cor.~13.21]{Pen2003} (\(k=0\), dimension \(d=2\), and \(p=2\)-norm distance function) which shows that the scaled minimum connectivity distance \(n\pi T^2 - \log(n)\) converges in distribution to a Gumbel distribution, and
  \((c)\) follows from the continuity of the Gumbel \cdf.

  Thus, for the given choice of \(\rn^2\):
  \begin{equation*}
    \limninfty \!\Prob{\Gnrn \!\in\! \Pconn} = \limninfty \!\Prob{T \!\leq\! \rn} = \erm^{-\erm^{-x}}. \IEEEQEDhereeqn
  \end{equation*}
\end{IEEEproof}

\subsection{%
  \lemref{rg-pgiant-threshold} (Reg.\ Threshold for \texorpdfstring{\(\Pgiant\)}{Pgiant} in \texorpdfstring{\(\Gnrn\)}{Gnrn})%
}\label{prf:rg-pgiant-threshold}

\begin{IEEEproof}
  Given \(\lambda_c \in (0,\infty)\), construct \(\rsn^2 = \lambda_c/n\).
  We show that:
  \begin{equation*}
    \rn^2 \sim x \rsn^2 \Rightarrow \limninfty \Prob{\Gnrn \in \Pgiant} = F(x),
  \end{equation*}
  for all points of continuity of \(F(x) \equiv \mathbf{1}\{x > 1\}\): \(\Rbb\setminus\{1\}\).

  Suppose \(\rn^2 \sim x\rsn^2\) with \(x \geq 0\).
  We have \(\rn^2 \sim \rho/n\) with \(\rho = x \lambda_c\).
  With \(h \in (0,1/x)\), there exists a single, bounded population cluster at level \(h\) equal to the unit square \(R_1 = [-1/2,1/2]^2 \subset \Rbb^2\).
  Let \(L_1\) be the normalized size of the largest component of \(\Gnrn\).
  By \cite[Thm.~11.9]{Pen2003}, we have that \(L_1\) converges in probability to \(I(R_1;\rho)\), since complete convergence implies convergence in probability:
  \begin{equation}\label{eq:penrose-thm11.9}
    \limninfty \Prob{\left|L_1/n - I(R_1;\rho)\right| > \eps} = 0, \forall \eps > 0,
  \end{equation}
  where \(I(R_1;\rho) = p_{\infty}(x\lambda_c)\) is the percolation probability under communication radius function \(\rsn^2 \sim  x\lambda_c / n\).

  Now, suppose that \(x < 1\).
  By definition, the percolation probability is zero, and:
  \begin{align}
    \limninfty &\Prob{\left|L_1 / n - I(R_1;\rho)\right| > \eps} = 0, \forall \eps > 0 \\
    \limninfty &\Prob{\left|L_1 / n\right| > \eps} = 0, \forall \eps > 0 \\
    \limninfty &\Prob{\Gnrn \in \Pgiant(\eps)} = 0, \forall \eps > 0 \\
    \limninfty &\Prob{\Gnrn \in \Pgiant} = 0.
  \end{align}

  Alternately, suppose that \(x > 1\).
  By definition, the percolation probability is positive, and:
  \begin{align}
    1 &=    \!\limninfty\! \Prob{\left| L_1 / n - I(R_1;\rho)\right| > \eps}, \forall \eps > 0 \\
      &\leq \!\limninfty\! \Prob{ L_1 / n \geq I(R_1;\rho) - \eps}, \forall \eps > 0 \\
      &=    \!\limninfty\! \Prob{ L_1 / n \geq p_{\infty}(x\lambda_c) - \eps}, \forall \eps \in (0,p_{\infty}(x\lambda_c)) \\
      &=    \!\limninfty\! \Prob{\Gnrn \!\in\! \Pgiant(\beta)}, \forall \beta \!\in\! (0,p_{\infty}(x\lambda_c)\!-\!\eps) \\
      &=    \!\limninfty\! \Prob{\Gnrn \in \Pgiant}.
  \end{align}
  By the squeeze theorem, we conclude that \(\limninfty \Prob{\Gnrn \in \Pgiant} = 1\).
\end{IEEEproof}

\subsection{%
  \thmref{rg-exclusion-plop-pgiant} (Mutual Excl.\ of \texorpdfstring{\(\Plop\)}{Plop} and \texorpdfstring{\(\Pgiant\)}{Pgiant} in \texorpdfstring{\(\Gnrn\)}{Gnrn})%
}\label{prf:rg-exclusion-plop-pgiant}

\begin{IEEEproof}
  By \lemref{rg-pgiant-threshold}, \(\rn^2 \sim c/n, c < \lambda_c\) implies that \(\Pgiant\) holds \aan.
  Therefore, \(\rn^2 \sim c/n, c \geq \lambda_c\) is a necessary condition for \(\Pgiant\) to hold \aas.
  Under this necessary condition, we see that \(\rn^2 \in \omega(1/n^{6/5})\) and by \corref{rg-plop-0-statement}, \(\Plop\) holds \aan.

  Thus, for \(\rn^2 \sim c/n, \forall c \leq \lambda_c\):
  \begin{equation}
    0 \leq \limninfty \Prob{\Pgiant \cap \Plop} \leq \limninfty \Prob{\Pgiant} = 0
  \end{equation}

  Alternately, for \(\rn^2 \sim c/n, \forall c > \lambda_c\):
  \begin{equation}
    0 \leq \limninfty \Prob{\Pgiant \cap \Plop} \leq \limninfty \Prob{\Plop} = 0
  \end{equation}

  In both cases, we can conclude that \(\limninfty \Prob{\Pgiant \cap \Plop} = 0\).
\end{IEEEproof}

\subsection{%
  \prpref{rg-sigma-prob-bounds} (\texorpdfstring{\(\sigma\)-LoP}{LoP Factor} Bounds in \texorpdfstring{\(\Gnrn\)}{Gnrn})
}\label{prf:rg-sigma-prob-bounds}

\begin{IEEEproof}
  Let \(G = \Gnrn\).
  \begin{align}
    \limninfty &\Prob{1/2 \leq \sigma(G) \leq 2/3} \stackrel{(a)}{=} \limninfty \Prob{\sigma(G) \leq 2/3} \\
    &\stackrel{(b)}{\geq} \limninfty \Prob{C_6 \subseteq G} \stackrel{(c)}{=} 1.
  \end{align}
  where
  \((a)\) the lower bound is always true by \lemref{sigma-bounds},
  \((b)\) the presence of \(C_6\) is sufficient for \(\sigma(G) \leq 2/3\) by \lemref{sigma-C6k}, and
  \((c)\) \(\rn^2\) is above the threshold for the appearance of \(C_6\), obtained from the argument of \prpref{rg-plop-upperbound} in \prfref{rg-plop-upperbound}.
\end{IEEEproof}

\subsection{%
  \thmref{rg-sigma-exp-bounds} (\texorpdfstring{\(\Expect{\sigma}\)}{Expected LoP Factor} Bounds in \texorpdfstring{\(\Gnrn\)}{Gnrn})
}\label{prf:rg-sigma-exp-bounds}

\begin{IEEEproof}
  Let \(G = \Gnrn\).
  The lower bound follows immediately from the support bound \(\sigma(G) \geq 1/2\) in \lemref{sigma-bounds}.
  The upper bound can be derived:
  \begin{align}
    \Expect{\sigma(G)}
    &= \Expect{\sigma(G) | K_6 \subseteq G} \Prob{K_6 \subseteq G} \nonumber \\
    &\qquad + \Expect{\sigma(G) | K_6 \nsubseteq G} \Prob{K_6 \nsubseteq G} \\
    &\stackrel{(a)}{\leq} \frac{2}{3} (1 - \Prob{K_6 \nsubseteq G}) + \Prob{K_6 \nsubseteq G} \\
    &= \frac{1}{3}\left(2 + \Prob{K_6 \nsubseteq G}\right),
  \end{align}
  where
  \((a)\) \(\sigma(G) \leq 2/3\) when \(K_6 \subseteq G\) (\ie, \(C_6 \subset K_6\) and the presence of \(C_6\) is sufficient for \(\sigma(G) \leq 2/3\)) and \(\sigma(G) \leq 1\) is always true.
  Finally, we may upper bound \(\limninfty \Prob{K_6 \nsubseteq G}\) with \(\exp(-(\pi c / 4)^{5} / 6! )\) by applying portions of the argument of \prpref{rg-plop-upperbound} in \prfref{rg-plop-upperbound}.
\end{IEEEproof}

\section*{Acknowledgment}

The authors would like to thank Dr.~A.~Sarkar at Western Washington University for helpful discussion and feedback in the preparation of this paper.
They are also grateful to the anonymous reviewers for their valuable comments and suggestions to improve the quality of this paper.


\bibliographystyle{IEEEtran}
\bibliography{IEEEabrv,refs}

\newpage 


\begin{IEEEbiography}[{\includegraphics[width=1in,height=1.25in,clip,keepaspectratio]{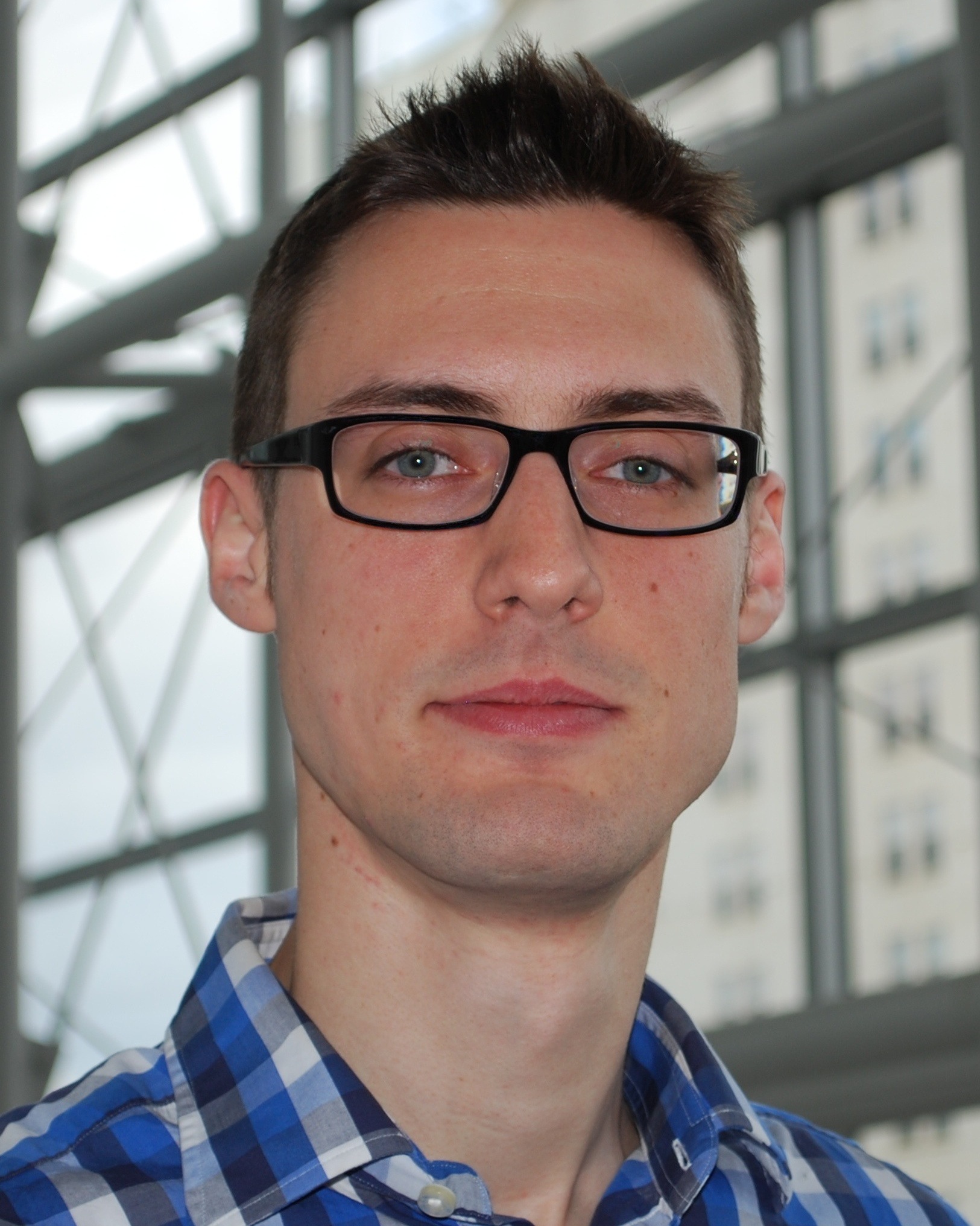}}]{Jeffrey Wildman}%
  (S'04-M'13) received the dual M.S./B.S. degree in electrical engineering from Drexel University, Philadelphia, PA, USA in 2009.
  He received the Ph.D. degree in electrical engineering from Drexel University in 2015.
  He has interned and visited with the Wideband Tactical Networking Group, MIT Lincoln Laboratory (2007, 2009); with the Network Operating Systems Technology Group, Cisco Systems, Inc. (2012); and most recently with the Center for Wireless Communications at the University of Oulu, Finland (2013).
\end{IEEEbiography}

\vspace*{-2.5\baselineskip}

\begin{IEEEbiography}[{\includegraphics[width=1in,height=1.25in,clip,keepaspectratio]{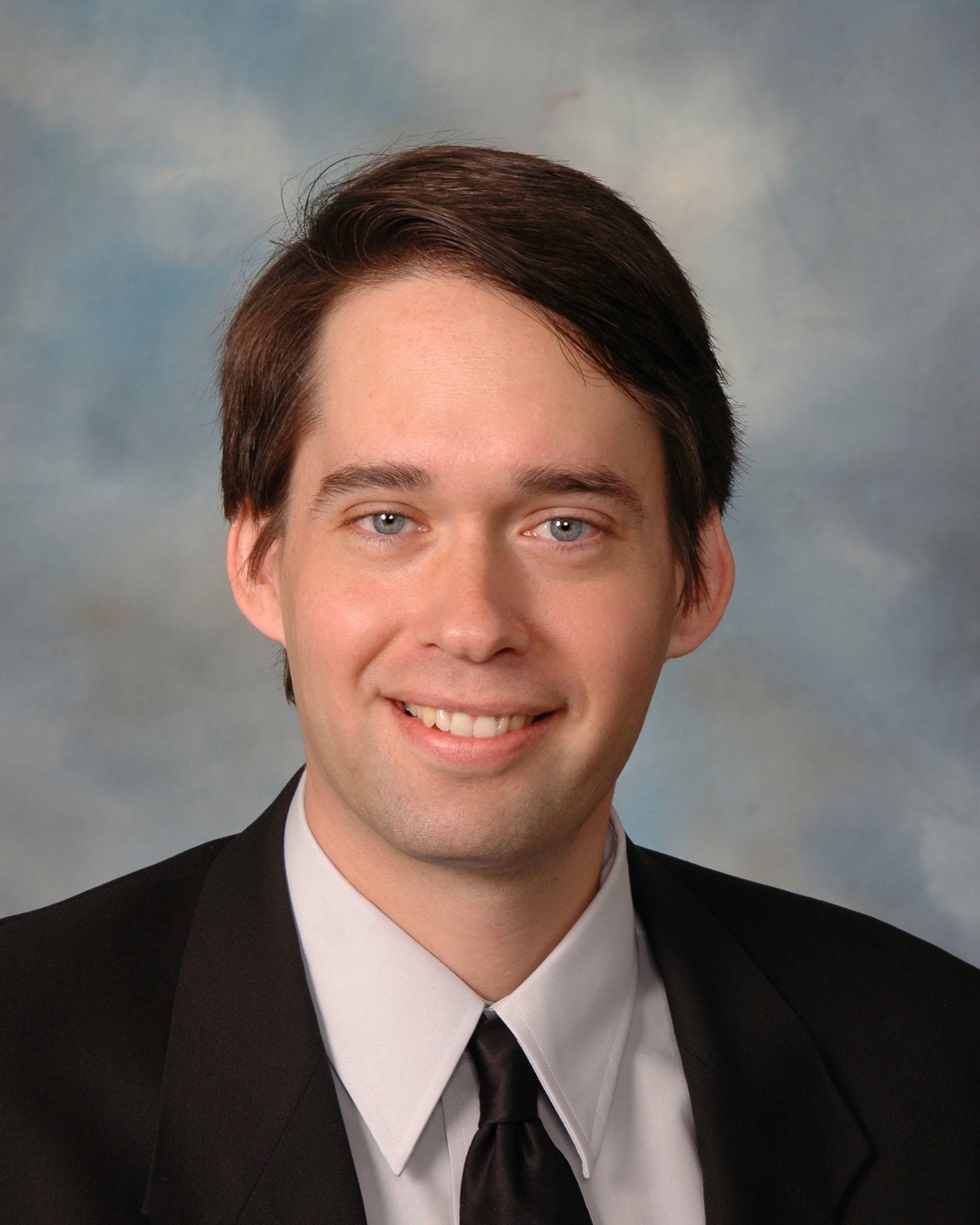}}]{Steven Weber}%
  (S'97-M'03-SM'11) received the B.S. degree in 1996 from Marquette University, Milwaukee, WI, USA, in 1996 and the M.S. and Ph.D. degrees from The University of Texas at Austin, TX, USA, in 1999 and 2003, respectively.
  He joined the Department of Electrical and Computer Engineering at Drexel University, Philadelphia, PA, USA, in 2003, where he is currently an Associate Professor.
  His research interests are centered around mathematical modeling of computer and communication networks, specifically streaming multimedia and ad hoc networks.
\end{IEEEbiography}

\vfill

\end{document}